\documentclass[usenatbib]{mn2e}
\usepackage[backref=none]{hyperref} %% To make sure backref is disabled
\usepackage[usenames,dvipsnames]{color}
\definecolor{MyDarkBlue}{rgb}{0,0.1,0.7}
\hypersetup{pdfborder={0 0 0},colorlinks,breaklinks=true,
  urlcolor={MyDarkBlue},citecolor={MyDarkBlue},linkcolor={MyDarkBlue} }

\usepackage{graphicx}
\usepackage{rotating}
\usepackage{amsmath,amssymb}
\usepackage{setspace}
\usepackage{longtable}

\newcommand{\apj}{ApJ}
\newcommand{\apjl}{ApJ}
\newcommand{\apjs}{ApJS}
\newcommand{\aj}{AJ}
\newcommand{\mnras}{MNRAS}
\newcommand{\nat}{Nature}

\newcommand{\aap}{A{\&}A}
\newcommand{\araa}{ARA{\&}A}

\newcommand{\apss}{ApSS}

\newcommand{\pasj}{PASJ}

\newcommand{\ssr}{Space Science Reviews}

\newcommand{\Msun}{\ensuremath{\mathrm{M}_{\bigodot}}}

\title[Constraints on the disc-magnetosphere interaction in accreting pulsar 4U 1626--67]
{Constraints on the disc-magnetosphere interaction in accreting pulsar 4U 1626--67}

\author[T{\"u}rko{\u g}lu et al.]
{M.\ Metehan T{\"u}rko{\u g}lu, G{\"o}k{\c c}e {\"O}zs{\"u}kan, M. Hakan Erkut,  K.\ Yavuz Ek\c{s}i  \\
  Istanbul Technical University,
  Faculty  of Science  and  Letters,  Physics Engineering  Department,
  34469,  Istanbul, Turkey \\
  \href{mailto:eksi@itu.edu.tr}{eksi@itu.edu.tr}}

\begin{document}
\date{}
\pagerange{\pageref{firstpage}--\pageref{lastpage}} \pubyear{2017}
\maketitle

\begin{abstract}
Using the spin and flux evolution of the accreting pulsar 4U 1626$-$67 across the 2008 torque reversal, we determine 
the fastness parameter dependence of the dimensionless torque acting on the pulsar. 
We find that the dimensionless torque is qualitatively 
different from the existing models: it is concave-up across the torque equilibrium 
whereas the existing torque models predict a concave-down (convex) relation with the fastness parameter. 
We show that the dimensionless torque has a cubic dependence on the fastness parameter 
near the torque equilibrium. 
We also find that the torque can not attain large values away from the equilibrium, either in the positive or the negative side, but saturates at limited values. 
The spin-down torque can attain a 2.5 times larger magnitude at the saturation limit than the spin-up torque.
From the evolution of the frequency of quasi-periodic oscillations of 4U 1626$-$67 across the torque reversal of 1990, we determine the critical fastness parameter corresponding to torque equilibrium to be $\omega_{\rm c} \simeq 0.75$ within the framework of the beat frequency model and boundary region model for reasonable values of the model parameters. 
We find that the disc magnetosphere interaction becomes unstable when the inner radius approaches the corotation radius as predicted by some models, though with a longer timescale. We also find that there is an unstable regime that is triggered when the fastness parameter is 0.8 times the critical fastness parameter ($\omega=0.6$ for $\omega_{\rm c} \simeq 0.75$) possibly associated with an instability observed in numerical simulations.
\end{abstract}


%%%%%%%%%%%%%%%%%%%%%%%%%%%%%%%%%%%%%%%%%%%%%%%%
\label{firstpage}

\section{INTRODUCTION}
\label{sec:intro}

X-ray pulsars are systems where a strongly magnetized neutron star disrupts the flow in the surrounding accretion disc
and channels it onto the surface of the neutron star \citep{nag89,bil+97}. The rate of gravitational potential energy released by accretion 
of matter onto the surface of the neutron star determines the X-ray luminosity as
\begin{equation}
L_{\mathrm{X}}=\frac{GM\dot{M}}{R},
\label{L_X}
\end{equation}
where $M$ is the mass, $R$ is the radius of the compact object, and $\dot{M}$ is the accretion rate.
Accreting matter carries angular momentum to the neutron star at the rate
\begin{equation}
N_0 = \sqrt{GMR_{\mathrm{in}}}\dot{M}  
\end{equation}
\citep{pri72}.  Here,  $R_{\mathrm{in}}$ is the inner radius of the disc determined by the balance of material and magnetic stresses \citep{gho79a} and scales with the Alfv{\'e}n radius \citep{els77,dav73} so that 
\begin{equation}
R_{\mathrm{in}}=\xi \left( \frac{\mu ^{2}}{\sqrt{2GM}\dot{M}}\right)^{2/7}
\label{R_in}
\end{equation}%
where $\mu$ is the magnetic moment of the neutron star and $\xi $ is a dimensionless number of order unity \citep[see e.g.][]{gho79b,wan96,rom+02}. Keplerian rotation in the disc matches the angular velocity of the star, $\Omega$, at the corotation radius
\begin{equation}
R_{\rm c}=\left( \frac{GM}{\Omega^2}\right)^{1/3}.
\label{corotation}
\end{equation}

A disc around an X-ray pulsar is highly ionized and has very high electrical conductivity.  As such, it is expected to show diamagnetic properties, i.e., that the magnetic field of the star be excluded 
from the disc due to the screening currents.
Accordingly, some models considered that the disc-magnetosphere interaction was limited by the innermost region of the disc where matter loads onto the magnetosphere \citep{sch78,aly80}. 
% Yet in order that the system displays X-ray pulsar properties matter should load onto the magnetosphere. except for the inner edge only for allowing loading of matter to the magnetosphere 
Discovery of pulsars showing spin-down episodes with no significant change in the X-ray luminosity led to the construction of the magnetically 
threaded disc model \citep{gho79a,gho79b}, in which the stellar field is assumed to penetrate the disc, via turbulent diffusion, reconnection and Kelvin-Helmholtz instabilities, 
over a wide radial range of the disc. The magnetic coupling between the poloidal stellar field, $B_z$, and the toroidal field on the upper surface of the disc, $B_{\phi }^{+}$, generated due to the rotational shear between the star and the disc
leads to the magnetic torque,
\begin{equation}
N_{\rm mag} = - \int_{R_{\rm in}}^{R_{\rm out}}B_{\phi }^{+}B_{z}r^{2}dr,
\label{N_mag}
\end{equation}
where $R_{\rm out}$ is the outer boundary of the coupling region.
The coupling between the stellar field and the  parts of the disc  rotating slower than the magnetosphere ($r>R_{\rm c}$) results in a spin-down torque on the star. Accordingly, accretion during spin-down is possible if this torque   dominates the sum of the spin-up torques i.e.\ the magnetic torque from the region $r<R_{\rm c}$ and the material torque $N_0$. The total torque $N=N_0 + N_{\rm mag}$ acting on the star can be written as
\begin{equation}
N \equiv n N_0, 
\label{torque}
\end{equation}%
which defines $n$, the dimensionless torque \citep{gho79a,gho79b}. In general, the dimensionless torque relies on the assumptions about the toroidal magnetic field in the disc and can be expressed in terms of the fastness parameter
\begin{equation}
\omega _{\ast }\equiv \frac{\Omega }{\Omega _{\mathrm{K}}\left( R_{\mathrm{in%
}}\right) }=\left( \frac{R_{\mathrm{in}}}{R_{\mathrm{c}}}\right) ^{3/2} 
\label{fastness}
\end{equation}
\citep{els77}.  The dimensionless torque depends on the fastness parameter, $n=n(\omega _{\ast })$. The total torque on the star vanishes at the critical fastness parameter, $\omega_{\rm c}$. For example, $\omega_{\rm c}=0.35$ according to the Ghosh-Lamb model \citep{gho79b}.

Further analysis revealed that the magnetic pressure of the toroidal field generated by the coupling of the stellar field with the accretion flow may become so high that the disc would be disrupted \citep{wan87}. 
This leads to possibilities like reconnection limiting the field in the disc \citep{wan95} or opening of the field lines \citep{aly90,uzd02,bar96,mat05}, which may 
remain open \citep{lov+95} or display alternating opening and reconnection episodes \citep{vanbal94,goo+97}.

The region in the disc that is penetrated by the stellar field lines and hence the torque acting on the pulsar remained a matter of debate investigated either analytically 
\citep[see e.g.][]{wan95,erk04,klu07,dai06,zha10,shu+94} or numerically \citep[see e.g.][]{hay+96,mil97,kol+02a,kol+02b,rom+02,rom+03,bra04,bes+08,zan09,zan13,kul13} and depends on the assumptions about the poorly constrained physics of turbulent magnetic  diffusivity and reconnection as well as on the grid resolution in the case of multi-dimensional numerical simulations \citep[see][for reviews]{uzd04,lai14,rom15}.

The usual theoretical approach in the study of disk magnetosphere interaction has been to investigate $n(\omega _{\ast })$, $\omega_{\rm c}$ and $\xi$
under certain assumptions about the processes limiting the toroidal field, configuration and topology of the magnetic fields, and prescriptions for magnetic diffusivity \citep[see e.g.][]{gho79b,wan95,klu07}. 
The purpose of the present paper 
is to construct the torque from observational data.
We do this ``reverse engineering'' in \S~\ref{sec:method}, using the spin and flux evolution of the accreting X-ray pulsar  4U~1626$-$67, and compare the observationally constructed torque with some existing torque models. In \S~\ref{sec:critical}, we determine, via the quasi-periodic oscillation (QPO) frequency evolution of 4U~1626$-$67,  the critical fastness, $\omega_{\rm c}$, corresponding to the torque equilibrium.
% In \S~\ref{sec:QPO} we investigate the relation between the Alfv\'en radius and the inner radius of the disc for X-ray pulsars from which both QPO frequency and cyclotron energy have been measured. 
% In \S~\ref{sec:pheno} we introduce a phenomenological model that could produce some properties of the determined torque.
Finally, in \S~\ref{sec:discussion}, we discuss the implications of our results 
 for understanding the disc-magnetosphere interaction.

%%%%%%%%%%%%%%%%%%%%%%%%%%%%%%%%%%%%%%%%%%%%%%%%
\section{An analysis of the accretion torques}
\label{sec:method}

In this section, we investigate the torque acting on X-ray pulsars.
In \S~\ref{subsec:1626} we apply the method to the X-ray pulsar 4U~1626--67  and  in \S~\ref{subsec:compare} we compare some of the existing torque models in the literature 
with the torque we constructed from available observational data.

\subsection{Method}
\label{subsec:method}

Assuming a beaming fraction of $b$, the flux received is
$F_{\mathrm{X}} = L_{\mathrm{X}}/4\pi b d^2$,
where $d$ is the distance of the source.
The mass flux then can be estimated from \autoref{L_X} as
\begin{equation}
 \dot{M}  = \frac{4\pi b d^2 F_{\mathrm{X}} R}{GM}  .
\label{mdot}
\end{equation}
Using this result in \autoref{R_in}, the
inner radius can be written as 
\begin{equation}
R_{\mathrm{in}} = \xi \left( \frac{ B^2 R^5 \sqrt{GM}}{16\sqrt{2}\pi b d^2 F_{\mathrm{X}}   }\right) ^{2/7},
\label{R_in2}
\end{equation}
where $B=2\mu /R^3$ is the magnetic field at the pole.

\begin{figure*}
\includegraphics[width=0.3\textwidth]{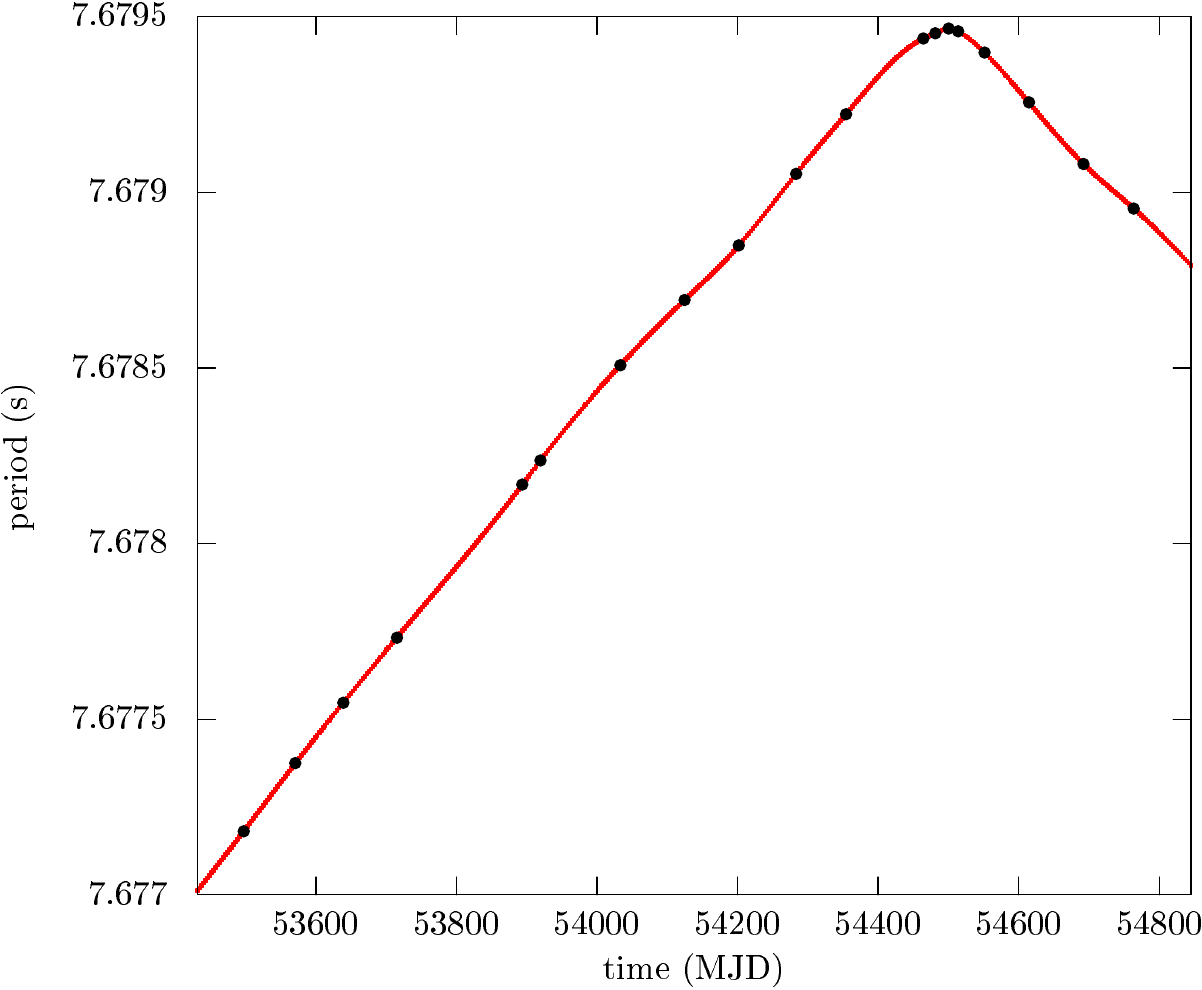}
\includegraphics[width=0.3\textwidth]{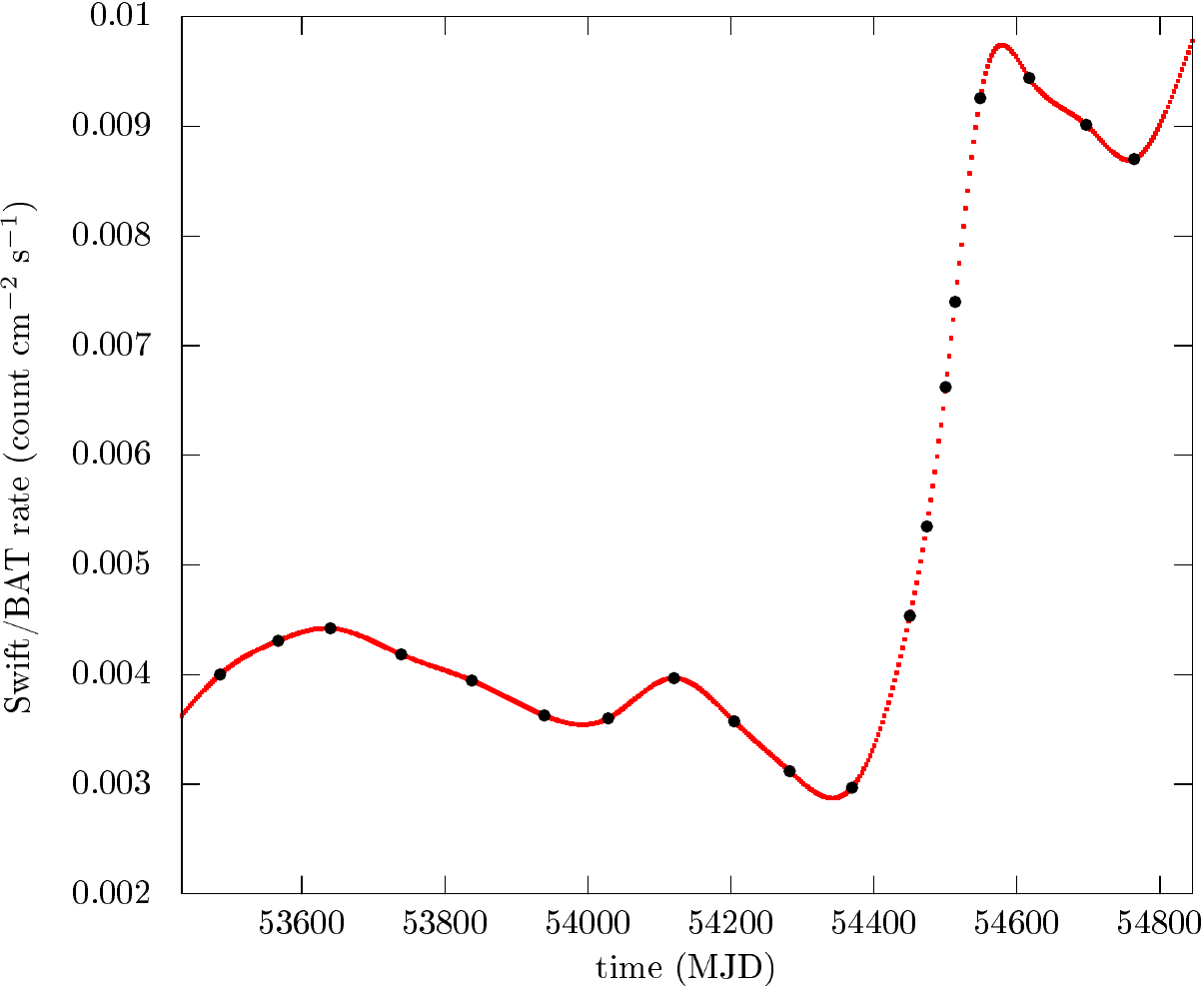}
\includegraphics[width=0.3\textwidth]{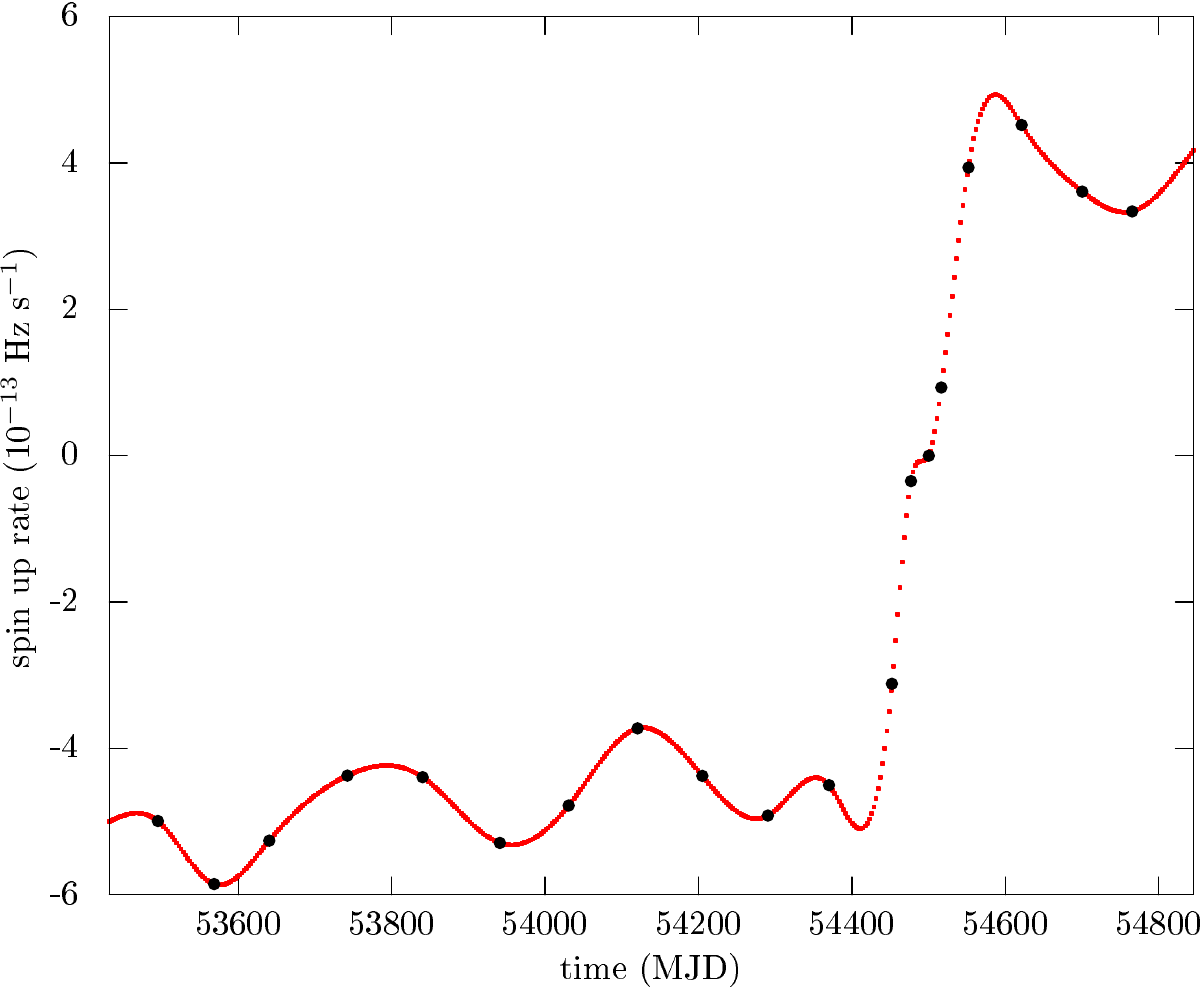}
\caption{Period, flux and spin-up rate ($\dot{\nu}$) evolution of 4U~1626$-$67. The observational data (black points) is taken from \citet{cam+10} and smoothed via interpolation into 600 data points (small red points).}
\label{fig:1626Period}
\end{figure*}

The fastness parameter defined in \autoref{fastness}, by referring to \autoref{R_in2} and 
\autoref{corotation}, can be written as 
\begin{equation}
\omega _{\ast }=\xi ^{3/2}\left( \frac{B^{2}R^{5}}{16\sqrt{2} {\mathrm \pi} b d^2}\right) ^{3/7}%
\frac{2{\mathrm \pi} }{\left( GM\right)^{2/7} F_{\mathrm{X}}^{3/7} P},
\label{fastness1}
\end{equation}
where $P =2\pi /\Omega$ is the rotation period.

The torque reversal occurs at a critical fastness parameter,
\begin{equation}
\omega _{\mathrm{c}}=\xi ^{3/2}\left( \frac{B^{2}R^{5}}{16{\mathrm \pi} \sqrt{2} b d^2}\right)
^{3/7}\frac{2{\mathrm \pi} }{\left( GM\right) ^{2/7}F_{\mathrm{c}}^{3/7}P_{\rm c}},
\label{fastness2}
\end{equation}
where $P_{\rm c}$ is the critical period, at which the torque reversal occurs and $F_{\rm c}$ is the corresponding flux.
Dividing \autoref{fastness1} with \autoref{fastness2}, we obtain
\begin{equation}
x\equiv \frac{\omega _{\ast }}{\omega _{c}}=\left( \frac{F_{\rm c}}{F_{\rm X}}%
\right) ^{3/7}\left( \frac{P_{\rm c}}{P}\right),
\label{xx}
\end{equation}
which is devoid of all unknown parameters like $B$, $d$, $M$ etc.
%It is a simple task to determine the time series of $x_i$ for this pulsar.
Note that implicit in the cancellation of $\xi$ in the above equation is the assumption that it is a constant. 
Many studies  \citep[e.g.][]{gho79b,wan96} find that the inner radius is independent of the rotation of the star, and yet it is possible that $\xi$ 
depends on the fastness parameter as well as the aspect ratio, $H/r$, of the disc. These dependences can be investigated through a detailed analysis of the non-keplerian boundary layer at the innermost disc \citep[see e.g.][]{gho79b}. We investigate this possibility in a subsequent paper.

The torque equation, $N=I\dot{\Omega}$, where  $I$ is the moment of inertia, by using \autoref{torque} and \autoref{R_in2},  can be written as 
\begin{equation}
-\dot{P}=n\omega _{\ast }^{1/3}\frac{4{\mathrm \pi} b d^2 R}{\left( 2{\mathrm \pi} \right) ^{4/3}\left(
GM\right) ^{1/3}I}F_{\mathrm{X}}P^{7/3}  ,
\label{pdot}
\end{equation}
or by using \autoref{xx} 
\begin{equation}
-\dot{P}=\frac{\omega _{\rm c}^{1/3}4{\mathrm \pi} b d^2 R F_{\rm c}P_{c}^{7/3}}{\left( 2{\mathrm \pi} \right)
^{4/3}\left( GM\right) ^{1/3}I}\frac{n}{x^2}.
\label{dotP}
\end{equation}
The dimensionless torque expanded into a Taylor series near torque equilibrium can be written as
\begin{equation}
n=n(1)+n'(1)(x-1)+\frac{1}{2!}n''(1)(x-1)^2 +\cdots. 
\end{equation}
As $n(1)=0$, the first term drops and when the system is close to equilibrium, terms higher than $O(1)$ are negligible so that the dimensionless torque can be written as  
\begin{equation}
n=n_{0}(1 - x), \qquad \mbox{for $x \simeq 1$} \label{neareq}
\end{equation}
where $n_0=-n'(1)$.
Using \autoref{neareq}, \autoref{dotP} can be written as
\begin{equation}
-\dot{P}=C \frac{1-x}{x^2}, \qquad \mbox{for $x \simeq 1$}
\label{CC1}
\end{equation}
where
\begin{equation}
C  \equiv \frac{\omega _{c}^{1/3}4{\mathrm \pi} b d^2 R F_{\rm c}P_{\rm c}^{7/3} n_0 }{\left( 2{\mathrm \pi} \right)
^{4/3}\left( GM \right) ^{1/3}I}
\label{CC2}
\end{equation}
is a constant. The value of $C$ can be determined from the slope of $-\dot{P}$ versus $(1-x)/x^2$ at torque equilibrium. One can then plug it back into \autoref{dotP} to find
\begin{equation}
\frac{n}{n_0} = -\frac{\dot{P} x^2}{C }.
\label{dimensionless_torque}
\end{equation}
As $\dot{P}$ is available as a time series, one can plot $n/n_0$ versus $x$. Eliminating $F_{\mathrm{c}}P_{\rm c}^{3/7}$ in \autoref{CC2} by referring to \autoref{fastness2}, one obtains
\begin{align}
C &= \frac{\pi}{2\sqrt{2}} \frac{n_0 \xi^{7/2}}{\omega_{\rm c}^2} \frac{R^6}{GMI}B^2 \nonumber \\
   &= 0.597 \times 10^{-11} \frac{n_0 \xi^{7/2}}{\omega_{\rm c}^2} \frac{ R_{6}^6}{M_{1.4}I_{45}}B_{12}^2,
\label{CC3} 
\end{align}
where $I_{45}=I/10^{45}~{\rm g~cm^2}$, $B_{12}=B/10^{12}~{\rm G}$, $R_{6}=R/10^{6}~{\rm cm}$, and $M_{1.4}=M/1.4~\Msun$.

\subsection{Application to 4U 1626--67}
\label{subsec:1626}

The investigation of the torque would best be applied to an accreting pulsar in a low mass X-ray binary (LMXB) system so that the object is accreting from a disc 
formed by Roche lobe overflow, rather than from a wind, so that there is no contribution to the X-ray luminosity and the torque from the wind of the companion, as would be the case in high mass X-ray binaries.
In order that the torque is measurable in spite of the inherent noise in the data, the neutron star should have a large magnetic field that can disrupt the disc at a large distance. 
Thus the analysis can not be applied to the millisecond pulsars in LMXBs as they have very low magnetic fields. 
Her~X$-$1 is an LMXB with large magnetic fields, yet periodic occultation of the source due to the precession of the warped inner disc results in a luminosity, which is not proportional to the accretion rate. Hence, the torque-luminosity correlation is not strong \citep{klo+09}.

The most suitable system for the investigation of the torque
is the persistent X-ray source 4U~1626$-$67  discovered by  UHURU \citep{gia+72}. 
It is a $7.66~{\rm s}$ X-ray pulsar \citep{rap+77} accreting from an
extremely low mass companion of $\sim 0.04~\Msun$ \citep{lev+88}. The orbital period of the system is 42~minutes \citep{mid+81,cha98}. The pulsar has a strong magnetic field in the $(2.4-6.3) \times 10^{12}~{\rm G}$ range
as inferred from cyclotron line \citep{orl+98}, or even stronger field of $8\times 10^{12}~{\rm G}$ as inferred from the energy dependence of the pulse profiles \citep{kii+86}. The X-ray luminosity of the source is $1.25 \times 10^{37}~{\rm erg~s^{-1}}$ \citep{whi+83} for an assumed distance of $6~{\rm kpc}$ as deduced from the spin-up rate \citep{pra+79}.  A more recent estimate of the distance is given as $9\pm 4~{\rm kpc}$ \citep{tak+16}. We stress, however, that the results of the present work are independent of the assumed distance to the source as we work with the dimensionless quantity $F_{\rm X}/F_{\rm c}$.

A unique property of this pulsar is the relatively smooth spin change
and two discrete torque reversals that occurred within about 40 yr.
The pulsar was spinning up with a characteristic time-scale of $\sim 5000~{\rm yr}$ when it was discovered in 1977 (stage~I). 
In 1990 it suffered a torque reversal and started to spin-down at about the same rate \citep[stage~II;][]{wil+93,cha+97a,krau+07}. 
After spinning down for about 18 years steadily, 
the source underwent another torque reversal in 2008 \citep[stage~III;][]{cam+10,jai+10}. Recently, Ghosh-Lamb model was used to estimate the distance as well as the mass and radius of the neutron star  \citep{tak+16}. In the present work, however, we do not presume a specific torque model, but compare the observed torque behaviour with the estimates of different models in the literature.

\begin{figure}
\includegraphics[width=0.49\textwidth]{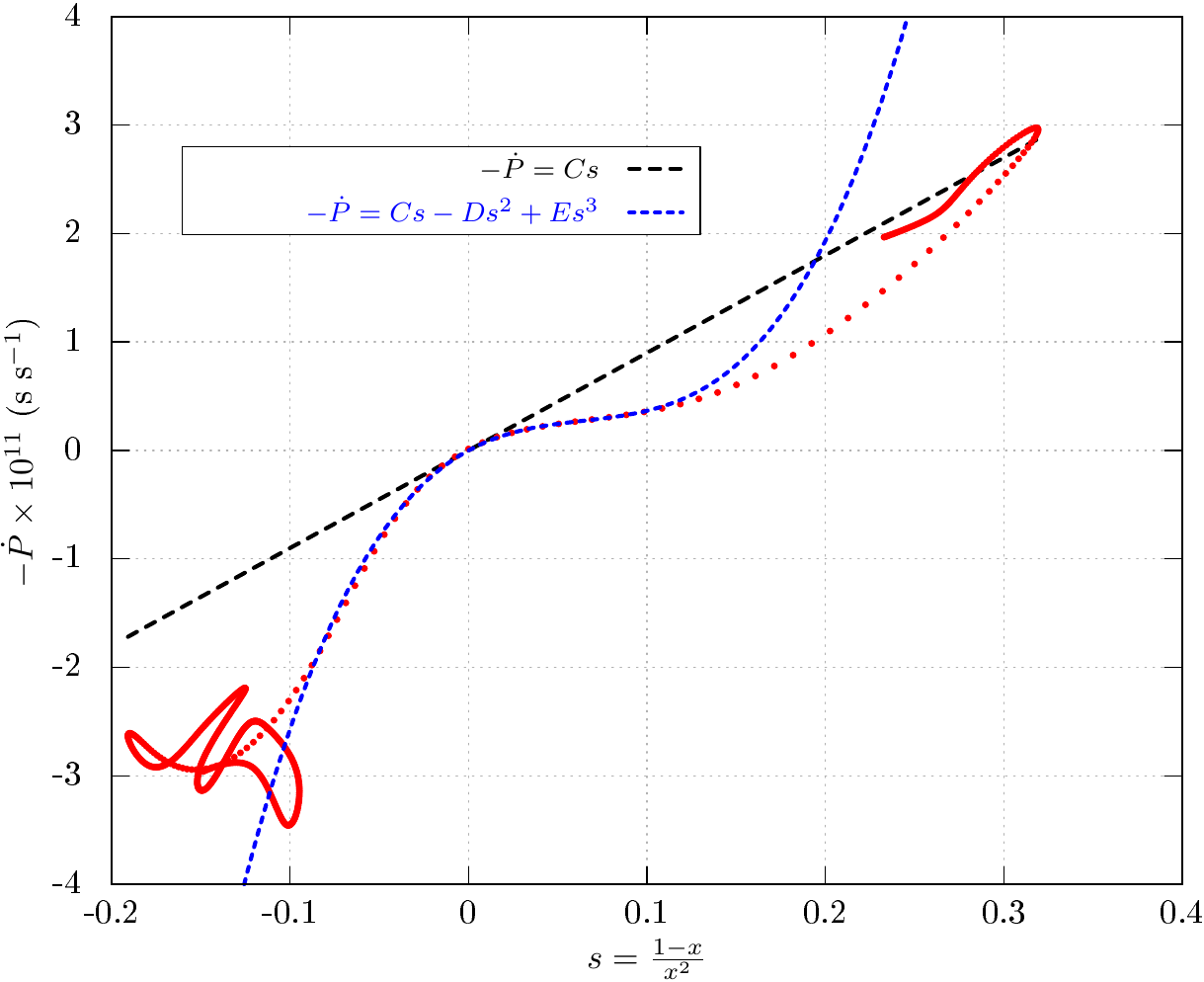}
\caption{$\dot{P}$ versus $s\equiv (1-x)/x^2$. The points (red in electronic version) are obtained by finding the time series of $x$ via \autoref{xx} for the X-ray pulsar 4U~1626--67.
The slope of the curve at the zero of the horizontal axis (i.e.\ at $x=1$) is shown with 
the long dashed line (black in the electronic version). The slope is found to be $9 \times 10^{-11}$ from the cubic polynomial fit shown with the short dashed (blue in the electronic version) curve.}
\label{fig:slope}
\end{figure}

\begin{figure}
\includegraphics[width=0.49\textwidth]{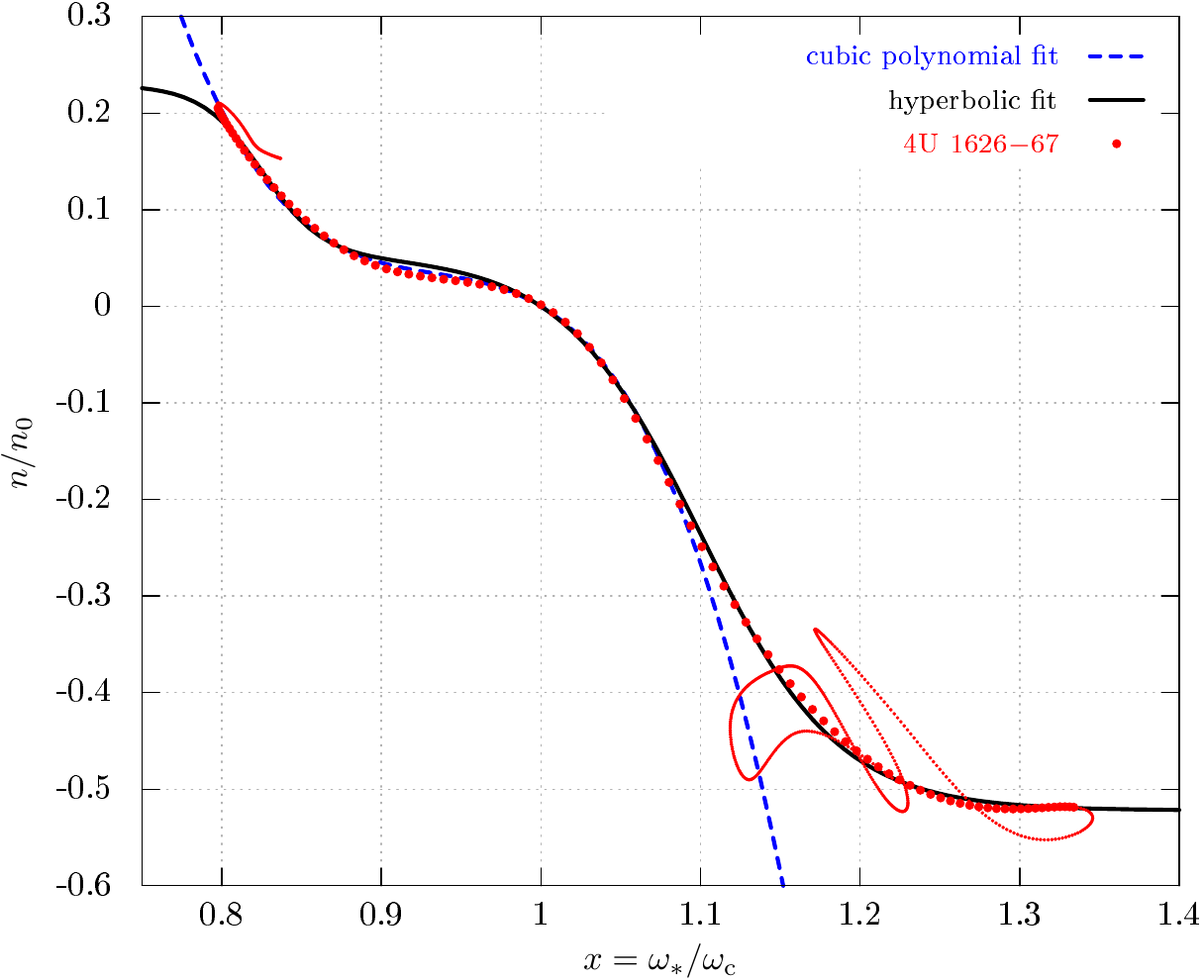}
\caption{The dimensionless torque $n$ scaled with the constant $n_0$ versus fastness parameter $\omega_{\ast}$ scaled with its critical value $\omega_{\rm c}$ corresponding to torque equilibrium.
The points (red in electronic version) are obtained by manipulating the observational period, flux and period derivative data of  the accreting pulsar 4U 1626$-$67 as described in the text.
The data in the erratic regime is shown with smaller points.
The long dashed curve (blue in the electronic version) 
is the cubic polynomial fit of the data in the range $x=0.8-1.1$ as given in Eqn.(\ref{polynomial}) with the best fit parameters given in Eqn.(\ref{ab}) and the solid black line is the trigonometric hyperbolic function given in Eqn.\ref{hyperbolic} with
the best fit parameters given in Eqn.~(\ref{fit}).}
\label{fig:torque1}
\end{figure}

In \autoref{fig:1626Period}, we show the period, flux, and frequency derivative evolution of the accreting pulsar 4U 1626--67 from the end of stage II to the beginning of stage III.
The data are taken from \citet{cam+10} and are smoothed via interpolation. It is seen that the torque reversal took place at around MJD~54500. Accordingly, the critical period 
is $P_{\rm c}=7.67946628~{\rm s}$ and the corresponding flux is
$F_{\rm c} = 0.00658~{\rm cnt~cm^{-2}~s^{-1} }$.  

Using \autoref{xx}, we have converted the data to a series of $x$ values and plotted $-\dot{P}$ versus $s=(1-x)/x^2$
as shown in \autoref{fig:slope}. In order to determine the value of $C$ given in \autoref{CC2}, we need to find the slope of the curve at $s=0$. 
In order to determine this accurately, we have fitted
the data in the range $s=-0.04$ to $-0.1$ with a cubic polynomial, $-\dot{P}=Cs-Ds^2+Es^3$, and found
\begin{align}
C &= (8.99579 \pm 0.1091)\times 10^{-11}\,   (1.213\%) \nonumber  \\
D &= (110.092 \pm 2.992)\times 10^{-11}\,    (2.718\%) \label{CDE} \\
E &= (567.082 \pm 32.73)\times 10^{-11}\,    (5.772\%) \nonumber . 
\end{align}
Hence, the value of the constant in \autoref{CC2} is determined to be $C \simeq 9 \times 10^{-11}$ being accurate to 1\%.

We have then used the value of $C$ to determine the dimensionless torque via \autoref{dimensionless_torque}.
The result is shown in \autoref{fig:torque1} with the use of filled circles (red in the electronic version).
Being a dimensionless function of a dimensionless parameter, one expects it to be universal for all X-ray pulsars accreting from a disc, and yet it may depend on another dimensionless parameter, the inclination angle between the rotation and magnetic axis of the pulsar.
We have fitted the data with a polynomial of the form
\begin{equation}
\frac{n}{n_0} = 1-x-a(x-1)^2 - b(x-1)^3
\label{polynomial}
\end{equation}
in the $x=0.8-1.1$ range, where we have obtained
\begin{equation}
a=11.0946 \pm 0.1787\,\,   (1.61\%)  \\
\end{equation}
and
\begin{equation}
b=55.7817 \pm 0.9935 \,\, (1.781\%). 
\label{ab}
\end{equation}
This is shown with dashed (blue in the electronic version) curve in \autoref{fig:torque1}.
This polynomial, although satisfactory for $x=0.8-1.1$, can not fit the torque for $x>1.1$, especially the tendency to saturate.
We have seen that the \textit{qualitative} form of the dimensionless torque all around the `smooth' regime, $x=0.8-1.34$, can be represented by a superposition of two hyperbolic functions as
\begin{align}
\frac{n}{n_0} =& \alpha_1 \left( \tanh \frac{x-x_1}{\delta_1} -\tanh \frac{1-x_1}{\delta_1} \right)  \nonumber \\
              &+ \alpha_2 \left( \tanh \frac{x-x_2}{\delta_2} - \tanh \frac{1-x_2}{\delta_2}\right),
\label{hyperbolic}              
\end{align}
where the constants (second terms in the parenthesis) guarantee the vanishing of the torque at $x=1$. The
best fit parameters are found as
\begin{align}
\alpha_1  &= -0.0886429  \pm 0.009559 \,    (10.78\%) \nonumber   \\
\delta_1  &= 0.0379771      \pm 0.004945 \,    (13.02\%) \nonumber  \\
x_1       &= 0.825271       \pm 0.004487 \,    (0.5437\%) \nonumber  \\
\alpha_2  &= -0.287013   \pm 0.002112 \,    (0.7359\%) \nonumber  \\
\delta_2  &= 0.0865664   \pm    0.001503 \,    (1.737\%) \nonumber  \\
x_2       & = 1.09982     \pm  0.0007801 \,   (0.07093\%).
\label{fit}
\end{align}
The function given in \autoref{hyperbolic} with these fit parameters is shown with the black solid line in \autoref{fig:torque1}. 
We note that although the hyperbolic 
function given in \autoref{hyperbolic} can represent a better portion of the curve, 
the cubic polynomial given in \autoref{polynomial} provides a better fit near the torque equilibrium.

\subsection{Comparison of the dimensionless torque with existing models}
\label{subsec:compare}

We would like to test some of the existing models in the literature with the result we inferred from observational data.
In order to compare our result with the Ghosh-Lamb model:
\begin{equation}
n_{\rm GL} = 1.39 \frac{1-\{ \omega_\ast [4.03(1-\omega_\ast)^{0.173} - 0.878] \}}{1-\omega_\ast},
\label{n_gl}
\end{equation}
for which $\omega_{\rm c}=0.35$ \citep{gho79a,gho79b}, we have calculated $n_0= -n_{\rm GL}'(1)=1.8839$ and have plotted
\begin{equation}
\frac{n_{\rm GL}}{n_0} = 0.737\frac{1-\{ \omega_{\rm c} x [4.03(1-\omega_{\rm c} x)^{0.173} - 0.878] \} }{1-\omega_{\rm c} x}.
\end{equation}
We have also compared our result with the dimensionless torques given by \citet{li96}, where the authors modify the torques given in an earlier work \citep{wan95}.
These torque models, listed in \autoref{table:torques}, are obtained under different assumptions, such as Alfv\'{e}n speed, turbulent diffusion, reconnection, buoyancy, and turbulence, about the physics limiting the growth of the toroidal field in the disc, while the toroidal field is generated
by the differential rotation between the magnetosphere and the disc \citep[see][for the details]{li96}.

\def\arraystretch{1.5}
\begin{table}
\caption{Different torque models presented by \citet{li96} which we use in \autoref{fig:torque2}.}
    \begin{tabular}{ | l | l | l | l |}
    \hline
    $n(x)=$ & physics limiting $B_{\phi}$ & $\omega_{\rm c}$ & $n_0$ \\ \hline
    $ 1+\frac{8}{9} \left( 1- \omega_{\rm c} x -\frac{(\omega_{\rm c} x)^{57/40}}{(1-\omega_{\rm c} x)^{1/2}} \right) $ & Alfv\'en speed & 0.76 & 4.4 \\ %\hline
    $ \frac{(5/3)-(7/3)\omega_{\rm c} x }{1-\omega_{\rm c} x}$ & turbulent diffusion & 5/7 & 35/6  \\ %\hline
    $ \frac{(5/3)-(7/3)\omega_{\rm c} x +(4/9)\omega^{2}_{\rm c} x^{2}}{(1-\omega_{\rm c} x)}$ & reconnection & 0.85 & 8.8 \\ %\hline
    $ 1+\frac{8}{9} \left( 1- \omega_{\rm c} x -\frac{(\omega_{\rm c} x)^{35/24}}{(1-\omega_{\rm c} x)^{1/2}} \right) $ & buoyancy & 0.76 &  4.3 \\ %\hline
    $ 1+\frac{20}{31}  \frac{1-(31/16)\omega_{\rm c} x}{1-\omega_{\rm c} x}  $ & turbulence & 0.73 & 6.1 \\ \hline
    \end{tabular}
    
\label{table:torques}
\end{table}

%   $\frac{n}{n_{0}}=\frac{6}{7} \frac{(7/6)-(4/3)\omega_{\rm c} x }{1-\omega_{\rm c} x}$ & turb.\ diff. & 0.875 & 9.333 \\ %\hline
%    $\frac{n}{n_{0}}=\frac{6}{7} \frac{(7/6)-(4/3)\omega_{\rm c} x +(1/9)\omega^{2}_{\rm c} x^{2}}{1-\omega_{\rm c} x}$ & rec. & 0.950 & 21.433 \\ %\hline

\begin{figure}
\includegraphics[width=0.47\textwidth]{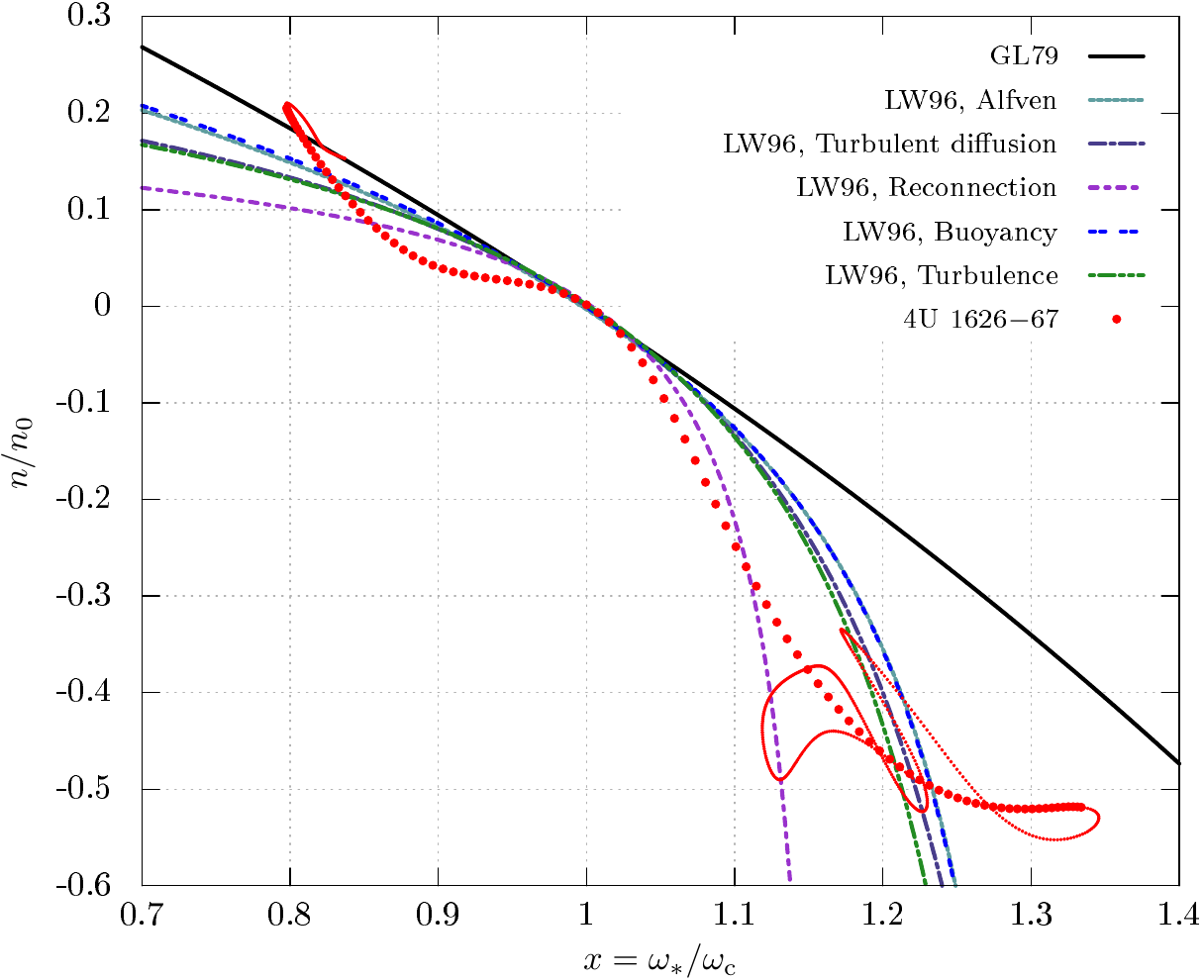}
\caption{Comparison of different torque models with the observationally constructed torque. The dimensionless torque $n$ is scaled to $n_0$ and the fastness parameter $\omega_{\ast}$ is scaled to its critical value $\omega_{\rm c}$ corresponding to torque equilibrium.
The points (red in electronic version) are obtained by manipulating the observational period, flux and period derivative data of  the accreting pulsar 4U 1626--67 as described in the text. Different curves correspond to different torque models, listed in \autoref{table:torques}, obtained under different assumptions about what limits the growth of the toroidal field (see text).}
\label{fig:torque2}
\end{figure}

In \autoref{fig:torque2}, we show these torque models together with the torque constructed from the spin-flux evolution of the accreting pulsar 4U 1626--67.
We see that the form of the torque obtained in this work is qualitatively different from the torque models
selected from the literature \citep{gho79b,li96}. The torques selected from the literature are concave-down (convex) across the torque equilibrium whereas 
the torque obtained in this work is concave-up. 
We are unaware of any model in the literature that predict such a property. We note, however, that the torque we have constructed is not purely observational, but a model dependent result as we employed certain assumptions: 
e.g.\ luminosity is assumed to be proportional to mass inflow rate in the disc as given in \autoref{L_X}, the inner radius is assumed to be proportional to the Alfv\'en radius i.e.\ $\xi$ in \autoref{R_in} is independent of the fastness parameter. We elaborate on these issues at the discussion section, but note here only that these reasonable assumptions are very commonly employed in the literature as is the case with the models we tested.

%%%%%%%%%%%%%%%%%%%%%%%%%%%%%%%%%%%%%%%%%%%%%%%%%%%%%%%%%%%%%%%%%%%%%%%%%%%%%%%%%%%%%%%%%%%%
%%%%%%%%%%%%%%%%%%%%%%%%%%%%%%%%%%%%%%%%%%%%%%%%%%%%%%%%%%%%%%%%%%%%%%%%%%%%%%%%%%%%%%%%%%%%
\section{Critical fastness parameter from the evolution of QPO frequency of 4U 1626--67}
\label{sec:critical}
%%%%%%%%%%%%%%%%%%%%%%%%%%%%%%%%%%%%%%%%%%%%%%%%%%%%%%%%%%%%%%%%%%%%%%%%%%%%%%%%%%%%%%%%%%%%
%%%%%%%%%%%%%%%%%%%%%%%%%%%%%%%%%%%%%%%%%%%%%%%%%%%%%%%%%%%%%%%%%%%%%%%%%%%%%%%%%%%%%%%%%%%%

\begin{figure*}
\includegraphics[width=0.49\textwidth]{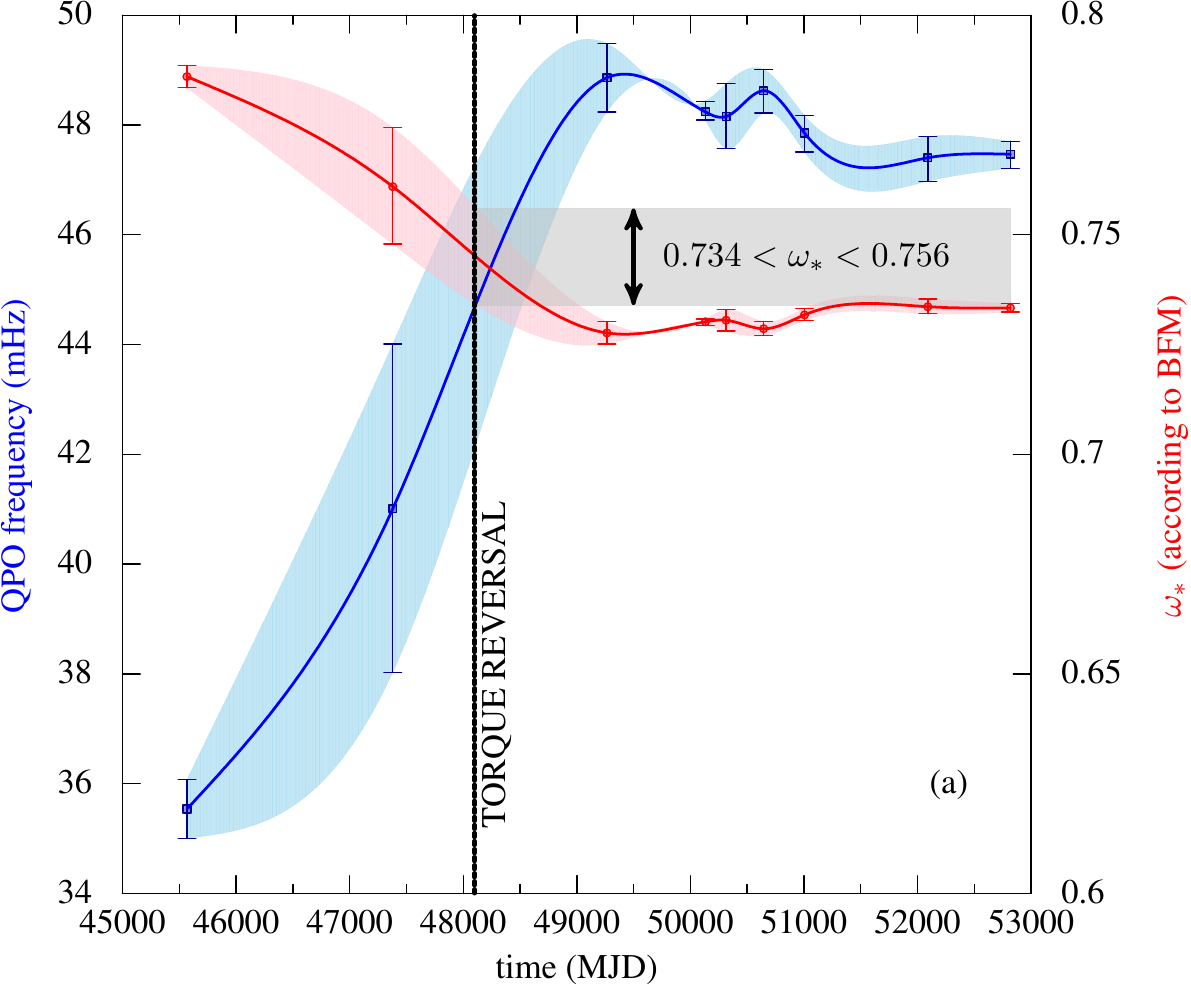}
\includegraphics[width=0.49\textwidth]{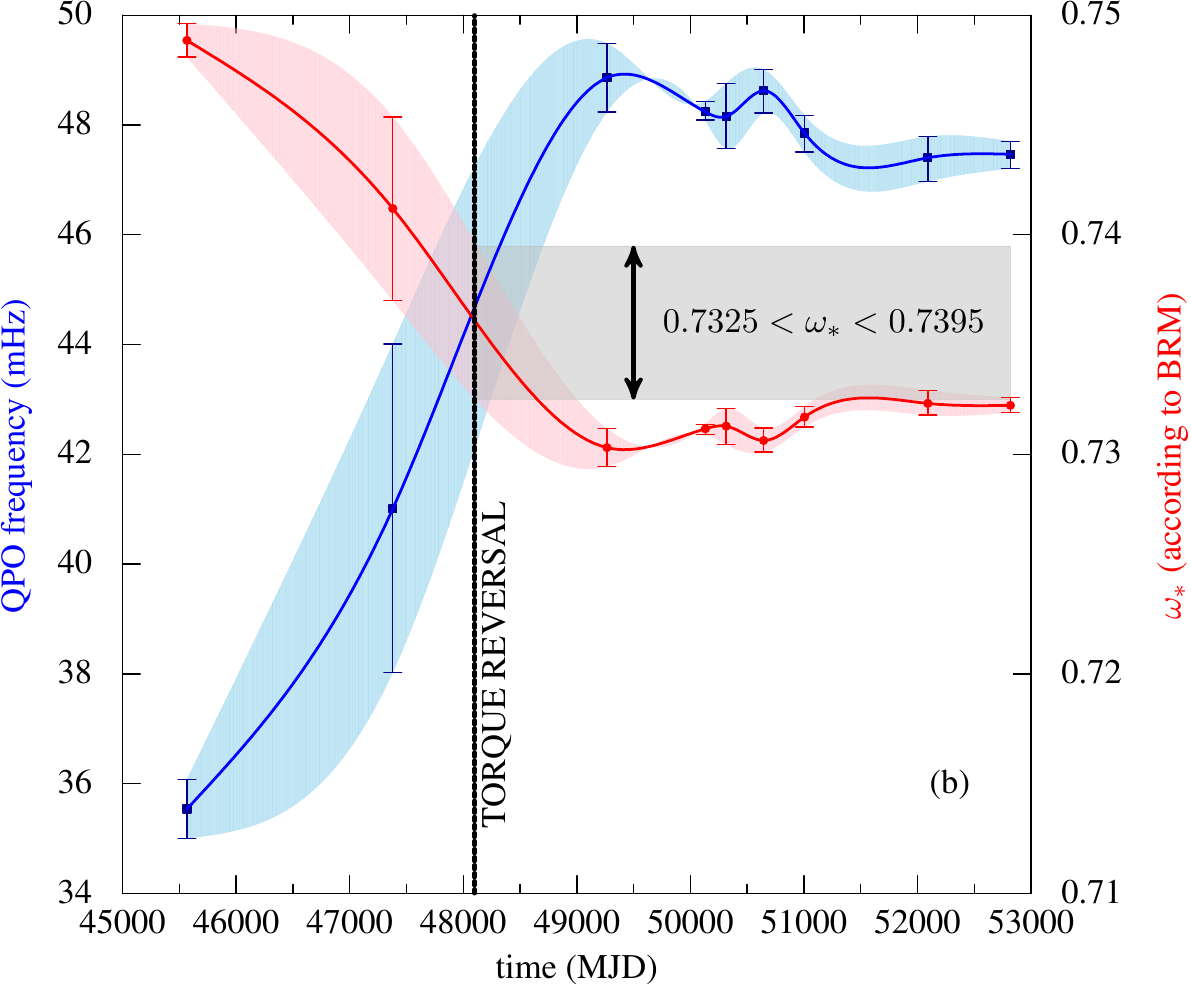}
\caption{The QPO frequency (left-y axis) evolution of 4U 1626--67 around the 1990 torque reversal (square points-blue in the electronic version) and evolution of the fastness parameter (right y-axis) of the object in the framework of BFM (left panel) and BRM (right panel). The fastness parameter for each QPO frequency is found by Eqn.(\ref{bfm}) for the BFM and by Eqn.(\ref{fsbrm} with $\delta=0.1$ and $m=3$) for the BRM (filled circles-red in the electronic version). 
The QPO frequency data is taken from the compilation of \citep{kau+08}. 
The solid lines correspond to the interpolated data values and the range of the errorbars are shaded. 
The dashed vertical line shows the torque reversal date that took place around June 1990. 
The width of the shaded rectangular region shown with the arrow correspond to the range of possible fastness parameter of the system at the date of the torque reversal i.e.\ $\omega_{\rm c} = 0.735-0.757$ for the BFM and $\omega_{\rm c} =0.7325 - 0.7395$ for the BRM.
}
\label{fig:qpo}
\end{figure*}

As a result of the magnetic coupling between the disc and the magnetosphere, several oscillatory modes in the inner disc might be excited in addition to the possible modulation of the X-ray emission from the neutron star by the inhomogeneities in the accretion flow. 
Whatever the specific mechanism of the variability is, the timescales of the representative features in the power spectrum, such as quasi-periodic oscillations (QPOs), are likely to be determined by the characteristic dynamical frequencies at a particular distance away from the compact object. 
The innermost disc radius, $R_{\rm in}$, sets in a natural length scale for the emergence of QPOs.
Observations of QPOs provide a probe of the position
of the inner disc radius in
accreting neutron stars \citep[see][for a review]{vanderklis00}. 

At a radial distance $r$ away from the neutron star, all dynamical frequencies can be expressed in terms of the Keplerian frequency,
\begin{equation}
\nu_{\mathrm{K}}\left(r\right)=\frac{1}{2\pi}\sqrt{\frac{GM}{r^3}}. \label{kepler}
\end{equation}

The frequency of QPOs as well as their quality observed from 4U 1626--67 changed in correlation with the torque state  \citep[see][for review]{kau+08}.
During stage~I, defined in \S~\ref{subsec:1626}, a weak and broad QPO at 40 mHz have been detected  with
Ginga \citep{shi+90} and ASCA \citep{ang+95}.  Later observations
made with BeppoSAX \citep{owe+97},
Rossi X-ray Timing Explorer \citep[RXTE;]{kom+98,cha98} and XMM-Newton \citep{krau+07} at stage~II show strong
QPOs at around 48 mHz with a slow frequency evolution with time. 
These 48 mHz QPOs disappeared at stage~III \citep{jai+10}. 
These results imply that QPOs in this system are weak or do not exist in the spin-up stages while they become prominent in the spin-down stages.

The torque reversal in 4U~1626--67 has been observed to be associated with a QPO feature detected at $\sim 42-47$~mHz. 
Knowing both the QPO frequency, $\nu_{\mathrm{QPO}}$, and the 
neutron-star spin frequency, $\nu_{\ast}$, it is possible to find the fastness parameter depending on a specific QPO model.
The evolution of the QPO frequency observed in 4U 1626--67 may thus be used to constrain the value of the critical fastness  parameter, $\omega_{\rm c}$, as the fastness parameter of the system at the date of the torque reversal.

We consider several QPO models to estimate the critical fastness of 4U~1626--67. 
First, we derive the expression for the fastness parameter using the model estimation of the QPO frequency. 
In each model, the QPO frequency is given as a function of the radial distance $r$ away from the compact object. 
For a magnetised accreting neutron star such as the one observed as an X-ray pulsar in 4U~1626--67, the magnetopause at $r=R_{\rm in}\simeq R_{\rm A}$ defines the relevant radius for the QPO frequency.

\subsection{Beat frequency model}

According to the beat frequency model \citep[BFM;][]{alp85,lam+85} QPOs arise due to
inhomogeneities in the inner accretion disc. The QPO frequency, $\nu_{\rm QPO}$, according to BFM, is the difference between 
the Keplerian frequency at the inner disc, $\nu_{\rm K}(R_{\rm in})$, and rotational frequency of the star, $\nu_{\ast}$:
\begin{equation}
 \nu_{\rm QPO} = \nu_{\rm K}(R_{\rm in}) - \nu_\ast .
\label{qpo1}
\end{equation}
Using \autoref{qpo1}, the fastness parameter, $\omega_\ast = \nu_{\ast}/ \nu_{\rm K}(R_{\rm in})$, in \autoref{fastness}, can be written as
\begin{equation}
\omega_\ast  = \frac{1}{1+\nu_{\rm QPO}/\nu_{\ast}}, \quad \mbox{in BFM}.
\label{bfm}
\end{equation}
We substitute $\nu_{\ast}=0.1305$~Hz and $\nu_{\mathrm{QPO}}\simeq 0.043$~Hz into \autoref{bfm} for the typical values of the spin and QPO frequencies during the torque reversal to find $\omega_c\simeq 0.75$ for the critical fastness of the source.
In \autoref{fig:qpo} the evolution of the QPO frequencies (left $y-$axis) are shown together with the evolution of the fastness parameter (right $y-$axis) calculated according to the BFM via \autoref{bfm}.
We have interpolated the data to estimate the values at the torque reversal date.
From the figure we see that the fastness parameter of the object, at the date of the torque reversal, is in the range $0.735-0.757$. We, thus, infer the critical fastness parameter for this system to be
\begin{equation}
\omega_{\rm c} = 0.746 \pm 0.011, \quad \mbox{in BFM}.
\label{w_c}
\end{equation}
This value is intermediate between two extreme values found in the literature:  $\omega_{\rm c}=0.5$ by \citet{gho79b} and $\omega_{\rm c} = 0.95$ by \citep{wan95}.
We emphasize that this is a model dependent result obtained within the framework of the BFM. Note that
the X-ray flux  gradually decreased during stage~I and during stage~II \citep{krau+07} while the QPO frequency increased, an observation against what BFM predicts \citep{zha10} \citep[see][for critics on BFM]{boz+09}. 
Note also that the precise value of the critical fastness 
parameter may depend on the inclination angle between rotation and magnetic axis and hence may have a different value for other systems.

\subsection{Keplerian frequency model}

The Keplerian frequency model \citep{Klis87} identifies the QPO frequency with the Keplerian frequency at the relevant radius. Using, $\nu_{\mathrm{QPO}}=\nu_{\mathrm{K}}\left(R_{\rm in}\right)$ with \autoref{fastness}, we find $\omega_{\ast}=\nu_{\ast}/\nu_{\rm QPO}$. The typical values of $\nu_{\ast}$ and $\nu_{\rm QPO}$, however, yield an unreasonably large value for the critical fastness, i.e., $\omega_c\simeq 3.04>1$. The relativistic models  \citep{SV98,SV99,SVM99,ABBK03} will not alter this estimate significantly because the inner radius of the disc is too large given that the neutron star is a slow rotator and strongly magnetized. The interpretation of the observed QPO frequencies as nodal precession frequency  $\nu_{\mathrm{nod}}=\nu_{\phi}-\nu_{\theta}$ on the other hand leads to extremely small values for the critical fastness parameter, $\omega_{\rm c} \sim 10^{-5}$ which is untenable.

\subsection{Magnetically driven precession model}

The diamagnetic feature of the accretion disc may induce a precessional torque on the inner disc matter. According to the magnetically driven precession model \citep{Lai99}, the misalignment of the magnetic dipole moment, $\mu$, with respect to the angular momentum of a ring of matter and the spin axis of the neutron star results in the precession of the ring of matter with frequency
\begin{equation}
\nu_{\mathrm{prec}}\left(r\right)=\frac{\mu^2 \left|\cos{\gamma}\right|\sin^2{\alpha}}{2\pi^3 r^7\Omega\left(r\right) \Sigma\left(r\right) D\left(r\right)}, \label{precf}
\end{equation}
where $\gamma$ is the angle between the neutron-star spin and the angular momentum of the ring of radius $r$, $\alpha$ is the angle between the neutron-star spin and the magnetic dipole moment, $\Omega$ is the orbital angular frequency, $\Sigma$ is the surface mass density, and $D=\left(2H/R_{\rm in}\right)^{1/2}$ with $H$ being the half-thickness of the disc. For $\nu_{\mathrm{QPO}}=\nu_{\mathrm{prec}}\left(R_{\rm in}\right)$, we employ the mass influx condition,
\begin{equation}
\Sigma \left(R_{\rm in}\right)=\frac{\dot{M}}{2\pi R_{\rm in}c_s\left(R_{\rm in}\right)}\left| \frac{c_s}{v_r}\right|_{R_{\rm in}} \label{mscnt}
\end{equation}
with $c_s\simeq 2\pi \nu_{\mathrm{K}}\left(Hr\right)^{1/2}$ being the sound speed in the magnetically dominated innermost disc region \citep{erk04}, the co-rotation with the magnetosphere, $\Omega\left(R_{\rm in}\right)=2\pi \nu_{\ast}$, and the estimation of $R_{\rm in}$ by \autoref{R_in} to write
\begin{equation}
\nu_{\mathrm{QPO}}=\frac{\sqrt{2}}{\pi}\left|\cos{\gamma}\right|\sin^2{\alpha}\left| \frac{v_r}{c_s}\right|_{R_{\rm in}}\frac{\nu^2_{\mathrm{K}}\left(R_{\rm in}\right)}{\nu_{\ast}}, \label{nqmp}
\end{equation}
which follows from \autoref{precf}. Using the definition in \autoref{fastness}, we obtain the fastness parameter as
\begin{equation}
\omega_{\ast}=\frac{2^{1/4}}{\sqrt{\pi}}\left|\cos{\gamma}\right|^{1/2}\sin{\alpha}\left| \frac{v_r}{c_s}\right|^{1/2}_{R_{\rm in}}\sqrt{\frac{\nu_{\ast}}{\nu_{\mathrm{QPO}}}}. \label{mpfs}
\end{equation}
Substituting for the typical values of $\nu_{\ast}$ and $\nu_{\mathrm{QPO}}$ during the torque reversal in 4U~1626--67, the critical fastness of the source can be estimated as
\begin{equation}
\omega_c\simeq 1.2\left|\cos{\gamma}\right|^{1/2}\sin{\alpha}\left| \frac{v_r}{c_s}\right|^{1/2}_{R_{\rm in}}. \label{cfmpr}
\end{equation}
Note that the value of $\omega_c$ depends on a few parameters such as $\alpha$, $\gamma$, and the ratio of the radial drift velocity, $v_r$, to the sound speed at the innermost disk radius.

\subsection{Boundary region model}

The global oscillatory modes in a non-Keplerian hydromagnetic boundary region may account for the QPOs observed in LMXBs. According to the boundary region model (BRM) of QPOs \citep{AP08,EPA08}, the frequency bands of the modes growing in amplitude are determined by the radial epicyclic frequency,
\begin{equation}
\nu_r=\nu_{\phi}\sqrt{4+2\frac{d\ln \nu_{\phi}}{d\ln r}}, \label{nrep}
\end{equation}
which is the highest dynamical frequency in a hydromagnetic boundary region with $\nu_{\phi}<\nu_{\mathrm{K}}$ \citep{AP08}.
For the low frequency QPOs, such as those observed during the torque reversal in 4U~1626--67, we consider the frequency bands, $\nu_r-m\nu_{\phi}$, where $m$ is the azimuthal wave number of the non-axisymmetric mode. We examine the possibility that $\nu_{\mathrm{QPO}}=\nu_r\left(R_{\rm in}\right)-m\nu_{\phi}\left(R_{\rm in}\right)$. At the magnetopause, $\nu_{\phi}\left(R_{\rm in}\right)=\nu_{\ast}$ for the matter co-rotating with the magnetosphere. Next, we employ \autoref{nrep} and estimate the radial epicyclic frequency at the same radius as
\begin{equation}
\nu_r\left(R_{\rm in}\right)\simeq \nu_{\ast}\sqrt{4+\frac{2}{\delta}\left[\frac{\nu_{\mathrm{K}}\left(R_{\rm in}\right)}{\nu_{\ast}}-1\right]}, \label{nurcl}
\end{equation}
where $\delta$ is the dimensionless width of the boundary region \citep{Erkut16}. Using \autoref{fastness}, it follows from \autoref{nurcl} that
\begin{equation}
\frac{\nu_{\mathrm{QPO}}}{\nu_{\ast}}\simeq \sqrt{4+\frac{2}{\delta}\left(\frac{1}{\omega_{\ast}}-1\right)}-m, 
\label{clfr}
\end{equation}
which we solve for the fastness parameter,
\begin{equation}
\omega_{\ast}\simeq \cfrac{1}{1+\cfrac{\delta}{2}\left[\left(m+\cfrac{\nu_{\mathrm{QPO}}}{\nu_{\ast}}\right)^2-4\right]}. \label{fsbrm}
\end{equation}
Note from \autoref{fsbrm} that $\omega_c<1$ is satisfied for the observed values of the spin and QPO frequencies during the torque reversal of 4U~1626--67, i.e., for $\nu_{\mathrm{QPO}}/\nu_{\ast}\approx 1/3$ only if $m\geq 2$. For $m=2$, $\omega_c\simeq 0.82$ if $\delta=0.3$ and $\omega_c\simeq 0.93$ if $\delta=0.1$. On the other hand, for $m=3$, $\omega_c\simeq 0.74$ if $\delta=0.1$ and $\omega_c\simeq 0.96$ if $\delta=0.01$. The observation that the fastness parameter decreases as the radial width of the boundary region increases has been revealed by earlier studies on the rotational dynamics of magnetically threaded discs \citep{erk04}.

\subsection{The erratic behaviour near boundaries of the fastness parameter}

We see from \autoref{fig:torque1} that the torque is multiply defined for $x \simeq 1.34$ ($\omega_\ast \simeq 1$ for $\omega_{\rm c}\simeq 0.75$)  and shows erratic dependence on the fastness parameter. As the system makes a transition from spin-down to spin-up, this erratic part is earlier in time and the system makes a transition to the smooth torque behaviour (larger data points) upon a transition across $\omega_\ast = 1$.  

This result is reminiscent of the interchange instability that is expected to set in when the inner radius is near the corotation \citep{sun77,spr93,dan10,dan12}, but the temporal resolution of the 
data we employed is much lower than the viscous timescale at the inner disc. Hence, we can not have resolved the instability considered by these authors with the present data. Yet we claim that 
the erratic behaviour of the torque is a longer timescale consequence of the instability as one of our assumptions, essential in our construction of the torque, breaks down should the instability set in. 

Implicit in our calculation of the torque is the assumption that the accretion rate determining the X-ray luminosity also determines the material stress and hence the inner radius of the disc.
When the inner radius is slightly beyond the corotation radius, it is expected that the centrifugal barrier does not allow all inflowing matter to reach the surface of the star. Yet the magnetosphere is 
incapable of ejecting the inflowing matter to escape velocity \citep{spr93}. As a result, matter accumulates outside the magnetosphere and the local accretion rate may be different than the accretion rate onto the star. The accretion rate determining the X-ray luminosity is no longer the same as the local accretion rate determining the material stress and the inner radius. So, an important assumption in our construction of the torque breaks down when the inner radius is slightly beyond the corotation radius. Hence, the fastness parameter associated with the erratic behaviour does not correspond to the true fastness parameter.

We see from \autoref{fig:torque1} that the torque is multiply valued also near $x \simeq 0.8$ ($\omega_\ast \simeq 0.6$ for $\omega_{\rm c}=0.75$). This behaviour is possibly the observational realization of the unstable chaotic regime that is expected to prevail for $0.45 < \omega_{\ast} < 0.6$ \citep{bli+16}. If this association is correct then the chaotic behaviour is due to the presence of several unstable tongues by which accretion proceeds.

\subsection{Magnetic field of the neutron star in 4U 1626--67}

The magnetic field of 4U 1626--67 inferred from the cyclotron feature at $\sim 37~{\rm keV}$ \citep{orl+98} is $B_{\rm c} = 3.2(1+z)\times 10^{12}~{\rm G}$ where $z$ is the gravitational redshift \citep[see also][]{cob+02}. This implies, for $z=0.3$, that $B_{\rm c}  \simeq 4 \times 10^{12}~{\rm G}$. An even larger value of $8\times 10^{12}~{\rm G}$ for the magnetic field was proposed by \citet{kii+86} from the energy dependence of the pulse profiles. More recently, \citet{dai+17} studied and modelled the broadband spectrum of the source finding evidence for the presence of a second harmonic at $\sim 61~{\rm keV}$.

From the value of $C=9\times 10^{-11}$ (see Eqn.(\ref{CDE})) and $\omega_{\rm c} \simeq 0.75$ (see Eqn.(\ref{w_c})) determined from the QPO frequency evolution, we can have an independent estimate of the magnetic field. 
Referring to \autoref{CC3} we obtain
\begin{equation}
B_{12} = 2.9 n_0^{-1/2}\xi^{-7/4} \left( \frac{\omega_{\rm c}}{0.75} \right) \frac{(M_{1.4}I_{45})^{1/2}}{R_{6}^3} .
\label{B2}
\end{equation}%
This is consistent with the value obtained from cyclotron lines as $n_0$ and $\xi$ are numbers of order unity. We would also like to note the possibility that the cyclotron feature may not be associated with the dipole field, but with a higher multipole near the surface \citep{alp12}.
If one assumes that the value of $B$ obtained from the cyclotron absorption feature truly reflects the surface dipole magnetic field of the star, then the equation above can be seen as an equation to constrain the combination $n_0\xi^{7/2} \simeq 0.53$.

\section{Discussion and Conclusion}
\label{sec:discussion}

Using available data of the X-ray pulsar 4U~1626--67 we have determined the dimensionless torque encompassing the essential physics of the disc-magnetosphere interaction.
Unfortunately the method can not be applied to any other known source. There are only a few persistent LMXB with large magnetic fields that show torque reversals.
In the case of Her X$-$1, the most well-known of these systems,  periodic occultation of the source due to the precession of the warped inner disc results in a luminosity not proportional to the accretion rate and as a consequence the torque luminosity correlation is not good \citep{klo+09}.

We have seen that the dimensionless torque near the equilibrium is qualitatively different than the predictions of the existing models. 
Across the torque equilibrium the torque is concave-up
in its dependence on the fastness parameter whereas the existing models \citep{gho79b,li96} predict concave-down (convex) relation. The torque has a cubic dependence near the torque equilibrium and the general qualitative behaviour can be represented by superposition of two hyperbolic functions and is shown to be very different than the existing analytical models.
These qualitative differences indicate either that an important ingredient of the disc-magnetosphere interaction is not captured in these models or that one or more of our assumptions in constructing the torque are not justified.

An assumption implicit in our construction of the torque is the proportionality of the inner radius with the Alfv\'en radius, $R_{\rm in} = \xi R_{\rm A}$, yet it is possible that $\xi$ is not a constant but depends on the fastness parameter. This condition will be relaxed in a future study, but we would like to mention that our comparison of the dimensionless torque with the existing torque models of \citet{gho79b} and \citet{li96} is self-consistent in the sense that these authors also find $\xi$ to be constant, independent of the fastness parameter.  

Another assumption that can be questioned is the proportionality of luminosity, and hence the flux received, to the accretion rate. There could be two ways this assumption may fail: It is possible that the inner disc is warped and obscures some of the X-ray flux as is the case in Her~X$-$1. This possibility is recently suggested by \citet{ber+15} relying on the pulse-phase dependence of the emission lines. Yet we note that the object shows a very smooth torque evolution near the torque equilibrium with a very satisfactory torque-flux correlation \citep{cam+10,cam+12} as opposed to the case in Her X-1. This is probably because we are seeing the system face on: We do not see the Doppler modulations, yet we do know the orbital period through optical modulation of the companion \citep{cha98}.

There is another way the linear relation between the accretion rate and the luminosity could break down. 
\citet{yi+97} suggested that the torque reversals from spin-up to spin-down could be a consequence of transition of the flow in the disc from a keplerian flow to a sub-keplerian flow. 
A torque reversal from spin-up to spin-down is then associated with a spectral transition from soft- to hard-state \citep{vau97,yi99}. 
Such spectral transitions, in the case of black holes, are associated with 
transition to hot accretion flow \citep[see][for a review]{yua14}    
in which the disc is optically thin/geometrically thick and energy is advected into the event horizon  \citep{ich77,nar94}. 
In the ADAF regime of black holes the disc luminosity is not linearly proportional to the accretion rate \citep{esi+97}. 
It is not clear how this picture could apply in the case of neutron stars where the X-ray luminosity has little contribution from the disc but arises simply by accretion onto a solid surface. 
If it leads to the possibility that the local accretion rate in the disc does not match to the accretion rate onto the neutron star an important ingredient in our assumptions break down.
%Our construction of the torque from the data  of  the accreting pulsar 4U 1626--67 presumes that the luminosity is linearly correlated with the accretion rate which also determines 
%the material torque and in turn the inner radius. 
While it is possible that the disc is in a state of hot accretion flow in the spin-down stage as supported by the harder spectrum, and assuming that the linear relation between luminosity and accretion rate is unjustified, this would only invalidate our results for the spin-down torque. 
The existing torque models are still challenged by the concave-up torque-fastness relation in the spin-up regime during which the spectrum is soft.

We have found, through the evolution of the frequency of QPOs in 4U~1626--67 \citep{kau+08}, that the critical fastness parameter corresponding to torque equilibrium is $0.746 \pm 0.011$ within the framework of the BFM \citep[but see][]{zha10}. This result is intermediate between the 0.5 of \citet{gho79b} and 0.95 of \citet{wan95}. Accordingly the erratic behaviour that appear at $x=\omega_\ast/\omega_{\rm c} \simeq 1.34$ in \autoref{fig:torque1} occurs around $\omega_\ast \simeq 1$ as shown in \autoref{fig:torque1}. This led us naturally to associate the erratic behaviour with the unsteady mass flux evolution that is expected to onset when the inner radius slightly exceeds the corotation radius \citep{sun77}. We emphasize that we do not claim to have observed the interchange instability \citep{spr93,dan10,dan12} that displays on the viscous time-scale of the inner disc which is shorter than the time resolution of our data.
Yet we claim that the erratic behaviour in the torque-fastness relation is a longer timescale consequence of it: an essential ingredient in our assumptions breaks down when the local accretion rate in the disc differs from the accretion rate onto the star.

We find that steady accretion onto the star proceeds for $x=0.8-1.34$ ($\omega_\ast = 0.6-1$ for $\omega_{\rm c}=0.75$). The erratic behaviour that onset when  fastness parameter drops below 
0.6 (see \autoref{fig:torque1}) could be associated with the so-called chaotic unstable regime observed in the numerical simulations of \citet{bli+16}. In this regime that prevails for $0.45 <\omega_\ast < 0.6$ the accretion proceeds with several tongues. Further analysis is required for associating our result with the numerical simulations which we defer for a subsequent paper.

%%%%%%%%%%%%%%%%%%%%%%%%%%%%%%%%%%%%%%%%%%%%%%%%
%%%%%%%%%%%%%%%%%%%%%%%%%%%%%%%%%%%%%%%%%%%%%%%%

\section*{Acknowledgements}
We thank Tolga G\"uver and M.\ Ali Alpar for useful comments.
KYE and MMT acknowledge support from The Scientific and Technological Council of TURKEY
(TUBITAK) with the project number 112T105. 
MHE acknowledges support from \.Istanbul Technical University for a postdoctoral fellowship.

%\bibliographystyle{mn2e}

%\bibliography{refs}

\begin{thebibliography}{}

\bibitem[\protect\citeauthoryear{{Abramowicz}, {Bulik}, {Bursa} \&
  {Klu{\'z}niak}}{{Abramowicz} et~al.}{2003}]{ABBK03}
{Abramowicz} M.~A.,  {Bulik} T.,  {Bursa} M.,    {Klu{\'z}niak} W.,  2003,
  \aap, 404, L21

\bibitem[\protect\citeauthoryear{{Alpar}}{{Alpar}}{2012}]{alp12}
{Alpar} M.~A.,  2012, \mnras, 423, 3768

\bibitem[\protect\citeauthoryear{{Alpar} \& {Psaltis}}{{Alpar} \&
  {Psaltis}}{2008}]{AP08}
{Alpar} M.~A.,  {Psaltis} D.,  2008, \mnras, 391, 1472

\bibitem[\protect\citeauthoryear{{Alpar} \& {Shaham}}{{Alpar} \&
  {Shaham}}{1985}]{alp85}
{Alpar} M.~A.,  {Shaham} J.,  1985, \nat, 316, 239

\bibitem[\protect\citeauthoryear{{Aly}}{{Aly}}{1980}]{aly80}
{Aly} J.~J.,  1980, \aap, 86, 192

\bibitem[\protect\citeauthoryear{{Aly} \& {Kuijpers}}{{Aly} \&
  {Kuijpers}}{1990}]{aly90}
{Aly} J.~J.,  {Kuijpers} J.,  1990, \aap, 227, 473

\bibitem[\protect\citeauthoryear{{Angelini}, {White}, {Nagase}, {Kallman},
  {Yoshida}, {Takeshima}, {Becker} \& {Paerels}}{{Angelini}
  et~al.}{1995}]{ang+95}
{Angelini} L.,  {White} N.~E.,  {Nagase} F.,  {Kallman} T.~R.,  {Yoshida} A.,
  {Takeshima} T.,  {Becker} C.,    {Paerels} F.,  1995, \apjl, 449, L41

\bibitem[\protect\citeauthoryear{{Bardou} \& {Heyvaerts}}{{Bardou} \&
  {Heyvaerts}}{1996}]{bar96}
{Bardou} A.,  {Heyvaerts} J.,  1996, \aap, 307, 1009

\bibitem[\protect\citeauthoryear{{Beri}, {Paul} \& {Dewangan}}{{Beri}
  et~al.}{2015}]{ber+15}
{Beri} A.,  {Paul} B.,    {Dewangan} G.~C.,  2015, ArXiv e-prints

\bibitem[\protect\citeauthoryear{{Bessolaz}, {Zanni}, {Ferreira}, {Keppens} \&
  {Bouvier}}{{Bessolaz} et~al.}{2008}]{bes+08}
{Bessolaz} N.,  {Zanni} C.,  {Ferreira} J.,  {Keppens} R.,    {Bouvier} J.,
  2008, \aap, 478, 155

\bibitem[\protect\citeauthoryear{{Bildsten}, {Chakrabarty}, {Chiu}, {Finger},
  {Koh}, {Nelson}, {Prince}, {Rubin}, {Scott}, {Stollberg}, {Vaughan}, {Wilson}
  \& {Wilson}}{{Bildsten} et~al.}{1997}]{bil+97}
{Bildsten} L.,  {Chakrabarty} D.,  {Chiu} J.,  {Finger} M.~H.,  {Koh} D.~T.,
  {Nelson} R.~W.,  {Prince} T.~A.,  {Rubin} B.~C.,  {Scott} D.~M.,  {Stollberg}
  M.,  {Vaughan} B.~A.,  {Wilson} C.~A.,    {Wilson} R.~B.,  1997, \apjs, 113,
  367

\bibitem[\protect\citeauthoryear{{Blinova}, {Romanova} \& {Lovelace}}{{Blinova}
  et~al.}{2016}]{bli+16}
{Blinova} A.~A.,  {Romanova} M.~M.,    {Lovelace} R.~V.~E.,  2016, \mnras, 459,
  2354

\bibitem[\protect\citeauthoryear{{Bozzo}, {Stella}, {Vietri} \&
  {Ghosh}}{{Bozzo} et~al.}{2009}]{boz+09}
{Bozzo} E.,  {Stella} L.,  {Vietri} M.,    {Ghosh} P.,  2009, \aap, 493, 809

\bibitem[\protect\citeauthoryear{{Camero-Arranz}, {Finger}, {Ikhsanov},
  {Wilson-Hodge} \& {Beklen}}{{Camero-Arranz} et~al.}{2010}]{cam+10}
{Camero-Arranz} A.,  {Finger} M.~H.,  {Ikhsanov} N.~R.,  {Wilson-Hodge} C.~A.,
    {Beklen} E.,  2010, \apj, 708, 1500

\bibitem[\protect\citeauthoryear{{Camero-Arranz}, {Pottschmidt}, {Finger},
  {Ikhsanov}, {Wilson-Hodge} \& {Marcu}}{{Camero-Arranz} et~al.}{2012}]{cam+12}
{Camero-Arranz} A.,  {Pottschmidt} K.,  {Finger} M.~H.,  {Ikhsanov} N.~R.,
  {Wilson-Hodge} C.~A.,    {Marcu} D.~M.,  2012, \aap, 546, A40

\bibitem[\protect\citeauthoryear{{Chakrabarty}}{{Chakrabarty}}{1998}]{cha98}
{Chakrabarty} D.,  1998, \apj, 492, 342

\bibitem[\protect\citeauthoryear{{Chakrabarty}, {Bildsten}, {Grunsfeld}, {Koh},
  {Prince}, {Vaughan}, {Finger}, {Scott} \& {Wilson}}{{Chakrabarty}
  et~al.}{1997}]{cha+97a}
{Chakrabarty} D.,  {Bildsten} L.,  {Grunsfeld} J.~M.,  {Koh} D.~T.,  {Prince}
  T.~A.,  {Vaughan} B.~A.,  {Finger} M.~H.,  {Scott} D.~M.,    {Wilson} R.~B.,
  1997, \apj, 474, 414

\bibitem[\protect\citeauthoryear{{Coburn}, {Heindl}, {Rothschild}, {Gruber},
  {Kreykenbohm}, {Wilms}, {Kretschmar} \& {Staubert}}{{Coburn}
  et~al.}{2002}]{cob+02}
{Coburn} W.,  {Heindl} W.~A.,  {Rothschild} R.~E.,  {Gruber} D.~E.,
  {Kreykenbohm} I.,  {Wilms} J.,  {Kretschmar} P.,    {Staubert} R.,  2002,
  \apj, 580, 394

\bibitem[\protect\citeauthoryear{{D'A{\`i}}, {Cusumano}, {Del Santo}, {La
  Parola} \& {Segreto}}{{D'A{\`i}} et~al.}{2017}]{dai+17}
{D'A{\`i}} A.,  {Cusumano} G.,  {Del Santo} M.,  {La Parola} V.,    {Segreto}
  A.,  2017, MNRAS accepted: DOI: https://doi.org/10.1093/mnras/stx1146

\bibitem[\protect\citeauthoryear{{Dai} \& {Li}}{{Dai} \& {Li}}{2006}]{dai06}
{Dai} H.-L.,  {Li} X.-D.,  2006, \aap, 451, 581

\bibitem[\protect\citeauthoryear{{D'Angelo} \& {Spruit}}{{D'Angelo} \&
  {Spruit}}{2010}]{dan10}
{D'Angelo} C.~R.,  {Spruit} H.~C.,  2010, \mnras, 406, 1208

\bibitem[\protect\citeauthoryear{{D'Angelo} \& {Spruit}}{{D'Angelo} \&
  {Spruit}}{2012}]{dan12}
{D'Angelo} C.~R.,  {Spruit} H.~C.,  2012, \mnras, 420, 416

\bibitem[\protect\citeauthoryear{{Davidson} \& {Ostriker}}{{Davidson} \&
  {Ostriker}}{1973}]{dav73}
{Davidson} K.,  {Ostriker} J.~P.,  1973, \apj, 179, 585

\bibitem[\protect\citeauthoryear{{Elsner} \& {Lamb}}{{Elsner} \&
  {Lamb}}{1977}]{els77}
{Elsner} R.~F.,  {Lamb} F.~K.,  1977, \apj, 215, 897

\bibitem[\protect\citeauthoryear{{Erkut} \& {Alpar}}{{Erkut} \&
  {Alpar}}{2004}]{erk04}
{Erkut} M.~H.,  {Alpar} M.~A.,  2004, \apj, 617, 461

\bibitem[\protect\citeauthoryear{{Erkut}, {Duran}, {{\c C}atmabacak} \& {{\c
  C}atmabacak}}{{Erkut} et~al.}{2016}]{Erkut16}
{Erkut} M.~H.,  {Duran} {\c S}.,  {{\c C}atmabacak} {\"O}.,    {{\c
  C}atmabacak} O.,  2016, \apj, 831, 25

\bibitem[\protect\citeauthoryear{{Erkut}, {Psaltis} \& {Alpar}}{{Erkut}
  et~al.}{2008}]{EPA08}
{Erkut} M.~H.,  {Psaltis} D.,    {Alpar} M.~A.,  2008, \apj, 687, 1220

\bibitem[\protect\citeauthoryear{{Esin}, {McClintock} \& {Narayan}}{{Esin}
  et~al.}{1997}]{esi+97}
{Esin} A.~A.,  {McClintock} J.~E.,    {Narayan} R.,  1997, \apj, 489, 865

\bibitem[\protect\citeauthoryear{{Ghosh} \& {Lamb}}{{Ghosh} \&
  {Lamb}}{1979a}]{gho79a}
{Ghosh} P.,  {Lamb} F.~K.,  1979a, \apj, 232, 259

\bibitem[\protect\citeauthoryear{{Ghosh} \& {Lamb}}{{Ghosh} \&
  {Lamb}}{1979b}]{gho79b}
{Ghosh} P.,  {Lamb} F.~K.,  1979b, \apj, 234, 296

\bibitem[\protect\citeauthoryear{{Giacconi}, {Murray}, {Gursky}, {Kellogg},
  {Schreier} \& {Tananbaum}}{{Giacconi} et~al.}{1972}]{gia+72}
{Giacconi} R.,  {Murray} S.,  {Gursky} H.,  {Kellogg} E.,  {Schreier} E.,
  {Tananbaum} H.,  1972, \apj, 178, 281

\bibitem[\protect\citeauthoryear{{Goodson}, {Winglee} \& {B{\"o}hm}}{{Goodson}
  et~al.}{1997}]{goo+97}
{Goodson} A.~P.,  {Winglee} R.~M.,    {B{\"o}hm} K.-H.,  1997, \apj, 489, 199

\bibitem[\protect\citeauthoryear{{Hayashi}, {Shibata} \& {Matsumoto}}{{Hayashi}
  et~al.}{1996}]{hay+96}
{Hayashi} M.~R.,  {Shibata} K.,    {Matsumoto} R.,  1996, \apjl, 468, L37

\bibitem[\protect\citeauthoryear{{Ichimaru}}{{Ichimaru}}{1977}]{ich77}
{Ichimaru} S.,  1977, \apj, 214, 840

\bibitem[\protect\citeauthoryear{{Jain}, {Paul} \& {Dutta}}{{Jain}
  et~al.}{2010}]{jai+10}
{Jain} C.,  {Paul} B.,    {Dutta} A.,  2010, \mnras, 403, 920

\bibitem[\protect\citeauthoryear{{Kaur}, {Paul}, {Kumar} \& {Sagar}}{{Kaur}
  et~al.}{2008}]{kau+08}
{Kaur} R.,  {Paul} B.,  {Kumar} B.,    {Sagar} R.,  2008, \apj, 676, 1184

\bibitem[\protect\citeauthoryear{{Kii}, {Hayakawa}, {Nagase}, {Ikegami} \&
  {Kawai}}{{Kii} et~al.}{1986}]{kii+86}
{Kii} T.,  {Hayakawa} S.,  {Nagase} F.,  {Ikegami} T.,    {Kawai} N.,  1986,
  \pasj, 38, 751

\bibitem[\protect\citeauthoryear{{Klochkov}, {Staubert}, {Postnov}, {Shakura}
  \& {Santangelo}}{{Klochkov} et~al.}{2009}]{klo+09}
{Klochkov} D.,  {Staubert} R.,  {Postnov} K.,  {Shakura} N.,    {Santangelo}
  A.,  2009, \aap, 506, 1261

\bibitem[\protect\citeauthoryear{{Klu{\'z}niak} \& {Rappaport}}{{Klu{\'z}niak}
  \& {Rappaport}}{2007}]{klu07}
{Klu{\'z}niak} W.,  {Rappaport} S.,  2007, \apj, 671, 1990

\bibitem[\protect\citeauthoryear{{Koldoba}, {Lovelace}, {Ustyugova} \&
  {Romanova}}{{Koldoba} et~al.}{2002a}]{kol+02a}
{Koldoba} A.~V.,  {Lovelace} R.~V.~E.,  {Ustyugova} G.~V.,    {Romanova} M.~M.,
   2002, \aj, 123, 2019

\bibitem[\protect\citeauthoryear{{Koldoba}, {Romanova}, {Ustyugova} \&
  {Lovelace}}{{Koldoba} et~al.}{2002b}]{kol+02b}
{Koldoba} A.~V.,  {Romanova} M.~M.,  {Ustyugova} G.~V.,    {Lovelace} R.~V.~E.,
   2002, \apjl, 576, L53

\bibitem[\protect\citeauthoryear{{Kommers}, {Chakrabarty} \& {Lewin}}{{Kommers}
  et~al.}{1998}]{kom+98}
{Kommers} J.~M.,  {Chakrabarty} D.,    {Lewin} W.~H.~G.,  1998, \apjl, 497, L33

\bibitem[\protect\citeauthoryear{{Krauss}, {Schulz}, {Chakrabarty}, {Juett} \&
  {Cottam}}{{Krauss} et~al.}{2007}]{krau+07}
{Krauss} M.~I.,  {Schulz} N.~S.,  {Chakrabarty} D.,  {Juett} A.~M.,    {Cottam}
  J.,  2007, \apj, 660, 605

\bibitem[\protect\citeauthoryear{{Kulkarni} \& {Romanova}}{{Kulkarni} \&
  {Romanova}}{2013}]{kul13}
{Kulkarni} A.~K.,  {Romanova} M.~M.,  2013, \mnras, 433, 3048

\bibitem[\protect\citeauthoryear{{Lai}}{{Lai}}{1999}]{Lai99}
{Lai} D.,  1999, \apj, 524, 1030

\bibitem[\protect\citeauthoryear{{Lai}}{{Lai}}{2014}]{lai14}
{Lai} D.,  2014, in European Physical Journal Web of Conferences Vol.~64 of
  European Physical Journal Web of Conferences, {Theory of Disk Accretion onto
  Magnetic Stars}.
p.~1001

\bibitem[\protect\citeauthoryear{{Lamb}, {Shibazaki}, {Alpar} \&
  {Shaham}}{{Lamb} et~al.}{1985}]{lam+85}
{Lamb} F.~K.,  {Shibazaki} N.,  {Alpar} M.~A.,    {Shaham} J.,  1985, \nat,
  317, 681

\bibitem[\protect\citeauthoryear{{Levine}, {Ma}, {McClintock}, {Rappaport},
  {van der Klis} \& {Verbunt}}{{Levine} et~al.}{1988}]{lev+88}
{Levine} A.,  {Ma} C.~P.,  {McClintock} J.,  {Rappaport} S.,  {van der Klis}
  M.,    {Verbunt} F.,  1988, \apj, 327, 732

\bibitem[\protect\citeauthoryear{{Li} \& {Wang}}{{Li} \& {Wang}}{1996}]{li96}
{Li} X.-D.,  {Wang} Z.-R.,  1996, \aap, 307, L5

\bibitem[\protect\citeauthoryear{{Lovelace}, {Romanova} \&
  {Bisnovatyi-Kogan}}{{Lovelace} et~al.}{1995}]{lov+95}
{Lovelace} R.~V.~E.,  {Romanova} M.~M.,    {Bisnovatyi-Kogan} G.~S.,  1995,
  \mnras, 275, 244

\bibitem[\protect\citeauthoryear{{Matt} \& {Pudritz}}{{Matt} \&
  {Pudritz}}{2005}]{mat05}
{Matt} S.,  {Pudritz} R.~E.,  2005, \mnras, 356, 167

\bibitem[\protect\citeauthoryear{{Middleditch}, {Mason}, {Nelson} \&
  {White}}{{Middleditch} et~al.}{1981}]{mid+81}
{Middleditch} J.,  {Mason} K.~O.,  {Nelson} J.~E.,    {White} N.~E.,  1981,
  \apj, 244, 1001

\bibitem[\protect\citeauthoryear{{Miller} \& {Stone}}{{Miller} \&
  {Stone}}{1997}]{mil97}
{Miller} K.~A.,  {Stone} J.~M.,  1997, \apj, 489, 890

\bibitem[\protect\citeauthoryear{{Nagase}}{{Nagase}}{1989}]{nag89}
{Nagase} F.,  1989, \pasj, 41, 1

\bibitem[\protect\citeauthoryear{{Narayan} \& {Yi}}{{Narayan} \&
  {Yi}}{1994}]{nar94}
{Narayan} R.,  {Yi} I.,  1994, \apjl, 428, L13

\bibitem[\protect\citeauthoryear{{Orlandini}, {Dal Fiume}, {Frontera}, {Del
  Sordo}, {Piraino}, {Santangelo}, {Segreto}, {Oosterbroek} \&
  {Parmar}}{{Orlandini} et~al.}{1998}]{orl+98}
{Orlandini} M.,  {Dal Fiume} D.,  {Frontera} F.,  {Del Sordo} S.,  {Piraino}
  S.,  {Santangelo} A.,  {Segreto} A.,  {Oosterbroek} T.,    {Parmar} A.~N.,
  1998, \apjl, 500, L163

\bibitem[\protect\citeauthoryear{{Owens}, {Oosterbroek} \& {Parmar}}{{Owens}
  et~al.}{1997}]{owe+97}
{Owens} A.,  {Oosterbroek} T.,    {Parmar} A.~N.,  1997, \aap, 324, L9

\bibitem[\protect\citeauthoryear{{Pravdo}, {White}, {Boldt}, {Holt},
  {Serlemitsos}, {Swank}, {Szymkowiak}, {Tuohy} \& {Garmire}}{{Pravdo}
  et~al.}{1979}]{pra+79}
{Pravdo} S.~H.,  {White} N.~E.,  {Boldt} E.~A.,  {Holt} S.~S.,  {Serlemitsos}
  P.~J.,  {Swank} J.~H.,  {Szymkowiak} A.~E.,  {Tuohy} I.,    {Garmire} G.,
  1979, \apj, 231, 912

\bibitem[\protect\citeauthoryear{{Pringle} \& {Rees}}{{Pringle} \&
  {Rees}}{1972}]{pri72}
{Pringle} J.~E.,  {Rees} M.~J.,  1972, \aap, 21, 1

\bibitem[\protect\citeauthoryear{{Rappaport}, {Markert}, {Li}, {Clark},
  {Jernigan} \& {McClintock}}{{Rappaport} et~al.}{1977}]{rap+77}
{Rappaport} S.,  {Markert} T.,  {Li} F.~K.,  {Clark} G.~W.,  {Jernigan} J.~G.,
    {McClintock} J.~E.,  1977, \apjl, 217, L29

\bibitem[\protect\citeauthoryear{{Romanova} \& {Owocki}}{{Romanova} \&
  {Owocki}}{2015}]{rom15}
{Romanova} M.~M.,  {Owocki} S.~P.,  2015, \ssr, 191, 339

\bibitem[\protect\citeauthoryear{{Romanova}, {Ustyugova}, {Koldoba} \&
  {Lovelace}}{{Romanova} et~al.}{2002}]{rom+02}
{Romanova} M.~M.,  {Ustyugova} G.~V.,  {Koldoba} A.~V.,    {Lovelace} R.~V.~E.,
   2002, \apj, 578, 420

\bibitem[\protect\citeauthoryear{{Romanova}, {Ustyugova}, {Koldoba}, {Wick} \&
  {Lovelace}}{{Romanova} et~al.}{2003}]{rom+03}
{Romanova} M.~M.,  {Ustyugova} G.~V.,  {Koldoba} A.~V.,  {Wick} J.~V.,
  {Lovelace} R.~V.~E.,  2003, \apj, 595, 1009

\bibitem[\protect\citeauthoryear{{Scharlemann}}{{Scharlemann}}{1978}]{sch78}
{Scharlemann} E.~T.,  1978, \apj, 219, 617

\bibitem[\protect\citeauthoryear{{Shinoda}, {Kii}, {Mitsuda}, {Nagase},
  {Tanaka}, {Makishima} \& {Shibazaki}}{{Shinoda} et~al.}{1990}]{shi+90}
{Shinoda} K.,  {Kii} T.,  {Mitsuda} K.,  {Nagase} F.,  {Tanaka} Y.,
  {Makishima} K.,    {Shibazaki} N.,  1990, \pasj, 42, L27

\bibitem[\protect\citeauthoryear{{Shu}, {Najita}, {Ostriker}, {Wilkin}, {Ruden}
  \& {Lizano}}{{Shu} et~al.}{1994}]{shu+94}
{Shu} F.,  {Najita} J.,  {Ostriker} E.,  {Wilkin} F.,  {Ruden} S.,    {Lizano}
  S.,  1994, \apj, 429, 781

\bibitem[\protect\citeauthoryear{{Spruit} \& {Taam}}{{Spruit} \&
  {Taam}}{1993}]{spr93}
{Spruit} H.~C.,  {Taam} R.~E.,  1993, \apj, 402, 593

\bibitem[\protect\citeauthoryear{{Stella} \& {Vietri}}{{Stella} \&
  {Vietri}}{1998}]{SV98}
{Stella} L.,  {Vietri} M.,  1998, \apjl, 492, L59

\bibitem[\protect\citeauthoryear{{Stella} \& {Vietri}}{{Stella} \&
  {Vietri}}{1999}]{SV99}
{Stella} L.,  {Vietri} M.,  1999, Physical Review Letters, 82, 17

\bibitem[\protect\citeauthoryear{{Stella}, {Vietri} \& {Morsink}}{{Stella}
  et~al.}{1999}]{SVM99}
{Stella} L.,  {Vietri} M.,    {Morsink} S.~M.,  1999, \apjl, 524, L63

\bibitem[\protect\citeauthoryear{{Sunyaev} \& {Shakura}}{{Sunyaev} \&
  {Shakura}}{1977}]{sun77}
{Sunyaev} R.~A.,  {Shakura} N.~I.,  1977, Pisma v Astronomicheskii Zhurnal, 3,
  262

\bibitem[\protect\citeauthoryear{{Takagi}, {Mihara}, {Sugizaki}, {Makishima} \&
  {Morii}}{{Takagi} et~al.}{2016}]{tak+16}
{Takagi} T.,  {Mihara} T.,  {Sugizaki} M.,  {Makishima} K.,    {Morii} M.,
  2016, \pasj, 68, S13

\bibitem[\protect\citeauthoryear{{Uzdensky}}{{Uzdensky}}{2002}]{uzd02}
{Uzdensky} D.~A.,  2002, \apj, 572, 432

\bibitem[\protect\citeauthoryear{{Uzdensky}}{{Uzdensky}}{2004}]{uzd04}
{Uzdensky} D.~A.,  2004, \apss, 292, 573

\bibitem[\protect\citeauthoryear{{van Ballegooijen}}{{van
  Ballegooijen}}{1994}]{vanbal94}
{van Ballegooijen} A.~A.,  1994, \ssr, 68, 299

\bibitem[\protect\citeauthoryear{{van der Klis}}{{van der
  Klis}}{2000}]{vanderklis00}
{van der Klis} M.,  2000, \araa, 38, 717

\bibitem[\protect\citeauthoryear{{van der Klis}, {Stella}, {White}, {Jansen} \&
  {Parmar}}{{van der Klis} et~al.}{1987}]{Klis87}
{van der Klis} M.,  {Stella} L.,  {White} N.,  {Jansen} F.,    {Parmar} A.~N.,
  1987, \apj, 316, 411

\bibitem[\protect\citeauthoryear{{Vaughan} \& {Kitamoto}}{{Vaughan} \&
  {Kitamoto}}{1997}]{vau97}
{Vaughan} B.~A.,  {Kitamoto} S.,  1997, ArXiv Astrophysics e-prints

\bibitem[\protect\citeauthoryear{{von Rekowski} \& {Brandenburg}}{{von
  Rekowski} \& {Brandenburg}}{2004}]{bra04}
{von Rekowski} B.,  {Brandenburg} A.,  2004, \aap, 420, 17

\bibitem[\protect\citeauthoryear{{Wang}}{{Wang}}{1987}]{wan87}
{Wang} Y.-M.,  1987, \aap, 183, 257

\bibitem[\protect\citeauthoryear{{Wang}}{{Wang}}{1995}]{wan95}
{Wang} Y.-M.,  1995, \apjl, 449, L153

\bibitem[\protect\citeauthoryear{{Wang}}{{Wang}}{1996}]{wan96}
{Wang} Y.-M.,  1996, \apjl, 465, L111

\bibitem[\protect\citeauthoryear{{White}, {Swank} \& {Holt}}{{White}
  et~al.}{1983}]{whi+83}
{White} N.~E.,  {Swank} J.~H.,    {Holt} S.~S.,  1983, \apj, 270, 711

\bibitem[\protect\citeauthoryear{{Wilson}, {Fishman}, {Finger}, {Pendleton},
  {Prince} \& {Chakrabarty}}{{Wilson} et~al.}{1993}]{wil+93}
{Wilson} R.~B.,  {Fishman} G.~J.,  {Finger} M.~H.,  {Pendleton} G.~N.,
  {Prince} T.~A.,    {Chakrabarty} D.,  1993, in {Friedlander} M.,  {Gehrels}
  N.,   {Macomb} D.~J.,  eds, American Institute of Physics Conference Series
  Vol.~280 of American Institute of Physics Conference Series, {Observations of
  isolated pulsars and disk-fed X-ray binaries.}.
pp 291--302

\bibitem[\protect\citeauthoryear{{Yi} \& {Vishniac}}{{Yi} \&
  {Vishniac}}{1999}]{yi99}
{Yi} I.,  {Vishniac} E.~T.,  1999, \apjl, 516, L87

\bibitem[\protect\citeauthoryear{{Yi}, {Wheeler} \& {Vishniac}}{{Yi}
  et~al.}{1997}]{yi+97}
{Yi} I.,  {Wheeler} J.~C.,    {Vishniac} E.~T.,  1997, \apjl, 481, L51

\bibitem[\protect\citeauthoryear{{Yuan} \& {Narayan}}{{Yuan} \&
  {Narayan}}{2014}]{yua14}
{Yuan} F.,  {Narayan} R.,  2014, \araa, 52, 529

\bibitem[\protect\citeauthoryear{{Zanni} \& {Ferreira}}{{Zanni} \&
  {Ferreira}}{2009}]{zan09}
{Zanni} C.,  {Ferreira} J.,  2009, \aap, 508, 1117

\bibitem[\protect\citeauthoryear{{Zanni} \& {Ferreira}}{{Zanni} \&
  {Ferreira}}{2013}]{zan13}
{Zanni} C.,  {Ferreira} J.,  2013, \aap, 550, A99

\bibitem[\protect\citeauthoryear{{Zhang} \& {Li}}{{Zhang} \&
  {Li}}{2010}]{zha10}
{Zhang} Z.,  {Li} X.-D.,  2010, \aap, 518, A19

\end{thebibliography}

\label{lastpage}

\end{document}